\documentclass[aps,prl,reprint,superscriptaddress]{revtex4-1}

\usepackage{graphicx}
\usepackage{caption}
\usepackage{subcaption}
\usepackage{lmodern}
\usepackage{amsmath}
\usepackage{amsfonts}
\usepackage{xfrac}
\usepackage{array}

\begin{document}

%\preprint{}

\title{Assessing complexity by means of maximum entropy models}


\author{Gregor Chliamovitch}
\email{Gregor.Chliamovitch@unige.ch}
\affiliation{Department of Computer Science, University of Geneva, Switzerland}
\affiliation{Department of Theoretical Physics, University of Geneva, Switzerland}
\author{Bastien Chopard}
\affiliation{Department of Computer Science, University of Geneva, Switzerland}
\author{Lino Velasquez}
\affiliation{Department of Computer Science, University of Geneva, Switzerland}
%\thanks{}


\date{\today}

\begin{abstract}
We discuss a characterization of complexity based on successive approximations of the probability density describing a system by means of maximum entropy methods, thereby quantifying the respective role played by different orders of interaction. This characterization is applied on simple cellular automata in order to put it in perspective with the usual notion of complexity for such systems based on Wolfram classes. The overlap is shown to be good, but not perfect. This suggests that complexity in the sense of Wolfram emerges as an intermediate regime of maximum entropy-based complexity, but also gives insights regarding the role of initial conditions in complexity-related issues.
\end{abstract}


\maketitle


\section{I. Introduction}

In the course of the last few decades, so many characterizations (sometimes at odds with each other) of complexity have appeared that it would be illusory to give a one-sentence statement encapsulating them all. Perhaps one of the most widely accepted such characterizations could be, in deliberately fuzzy terms, that ``complexity arises when a system is more than the collection of its parts'' \cite{Bar-Yam2004b}. It is nonetheless far from easy to understand what ``being more than one's parts'' really means. A similar idea is conveyed by the statement that ``a complex system cannot be fully understood by looking separately at its elements'' which is barely more transparent but, as we shall see, paves the way to quantitative interpretations (see also \cite{Bar-Yam2004a}). As an historical aside, let us note that besides being the most popular, this notion of complexity also turns out to be one of the oldest since it can be traced back (in somewhat different terms) to Aristotle's writings \cite{Aristotle}. For convenience, we shall refer to this definition of complexity as \textit{Aristotle's complexity} or simply \textit{A-complexity}.

In recent years, it has been proposed \cite{Schneidman2003, Schneidman2006, Ay2011} to give a mathematical meaning to these abstract principles in a probabilistic framework by means of \textit{maximum entropy (ME) models}. The ME approach provides a conceptually simple way to build generic models based on observational constraints. Following this line of reasoning, Aristotle's principle could be reformulated by asserting that a system is complex when its ME approximation built on the knowledge of small subparts provides a poor approximation of the system as a whole.

ME models come in different versions, depending on the observational constraints retained for consideration. Recently, impressive successes have been obtained in the study of neural networks by considering ME models based on constrained two-points correlations and firing rates \cite{Schneidman2006}. However, little emphasis has been put on quantifying the limitations of this approach, and correlations are certainly not the only nor the most general quantity worth being considered. On another side, it has been suggested building ME reconstructions on the knowledge of marginals up to a certain order \cite{Schneidman2003, Ay2011}. While this second approach is more general and certainly more in accordance with the spirit of information theory, it has the drawback that the models so generated are more difficult to investigate analytically than their correlation-based counterparts, often resulting in fairly abstract statements and conclusions.

The purpose of the present work is to display the ME method ``at work'' by carrying through a numerical investigation of these marginal-based models. To this end we shall turn our attention to so-called \textit{one-dimensional elementary cellular automata (ECA)}, since for this kind of system a notion of complexity is well established and may therefore serve as a benchmark. More precisely, ECA have long since raised a huge interest due to the fact that they may be classified in four classes ranging from trivial to complex behaviour, in a sense to be discussed further later on. While this classification scheme (essentially due to Wolfram \cite{Wolfram1983}, which is why we shall refer to this notion of complexity as \textit{W-complexity}) is the most famous one, many others have been proposed over time (see \cite{Martinez2013} and references therein). Some attention will also be devoted in this work to \textit{Langton's parameter} \cite{Langton1990}, which, rather than a classification scheme, provides a parametrization of the CA space.

Despite innumerable attempts to apply usual information-theoretic tools to the study of cellular automata, inspired by the belief that complexity should, in the end, have something to do with information, we are not aware of a situation where these tools provide convincingly original and deep insights into these systems. It will turn out that ME tools give quite significant results in this context, namely by tying a link between W-complexity and the dependence of A-complexity on the size of the system, somewhat at odds with the idea that systems become more and more complex when they grow larger due to more and more room left for building synergistic interactions.

The outline of this paper is as follows. We start with a review of maximum entropy methods, and then show how these tools may be used to cast Aristotle's intuition into a proper mathematical scheme. We continue with a reminder on ECA and discuss how to implement these systems in a way that fits our purpose, after what our results are presented.

\section{II. Maximum entropy approximations}

The philosophy underlying ME models is so to speak opposite to the constructive one, where a model is built and tuned to match observed properties (\textit{top-down}). In the ME approach on the other side, we proceed by seeking the least structured model compatible with a given set of observations (\textit{bottom-up}). This is done by noting that the most general (that is the least structured) probability density is the one which has the largest entropy while still satisfying the observational constraints, which are usually provided by some set of observables $f_k$ ($k = 1, 2, ..., K$) the average values of which are known, $\langle f_k \rangle =\mu_k$. Note that a drawback of this procedure lies in the fact that it only yields a probability density. An energy function can be deduced by analogy however, but this important point will not be of much concern in the present paper.

Assume we look for a probability distribution $p$ on a set of $N$ variables $X_1, ..., X_N$ (collectively denoted by $\mathbf X$), such that $H(\mathbf X)=-\sum_{\mathbf x} p(\mathbf x) \ln p(\mathbf x)$ is maximal, while making sure that the constraints $\langle f_k \rangle := \sum_{\mathbf x} f_k(\mathbf x) p(\mathbf x) =\mu_k$ and $\sum_{\mathbf x} p(\mathbf x) = 1$ are enforced. Using Lagrange's method, the quantity we have to equate to zero is

\begin{widetext}

\begin{equation}
\frac{\partial}{\partial p (\mathbf y )} \left( -\sum_{\mathbf x} p(\mathbf x) \ln p(\mathbf x)  + \lambda_0 \left( \sum_{\mathbf x} p(\mathbf x) - 1  \right) + \sum_{k=1}^K \lambda_k \left( \sum_{\mathbf x} f_k(\mathbf x) p(\mathbf x) - \mu_k  \right) \right) ,
\end{equation}

\end{widetext}
where the $\lambda$'s are the multipliers.

Some straightforward algebra yields that the sought-after distribution may be written as

\begin{equation}
p(\mathbf x) = \frac{1}{Z} \exp \left( \sum_{k=1}^K \lambda_k f_k(\mathbf x) \right) ,
\end{equation}
where the multipliers have to be chosen to match the constraints. Dividing by the partition function $Z :=\sum_{\mathbf x} \exp \left( \sum_{k=1}^K \lambda_k f_k(\mathbf x) \right)$ ensures that $p$ is properly normalized. For instance in the most elementary case where no constraint besides normalization is imposed, the ME distribution is nothing but the uniform distribution (if the support of $p$ is unbounded we have to impose finite mean and variance in order to recover a meaningful distribution, which turns out to be a Gaussian).

While in the present context we use the ME approach as a tool to investigate how well the reconstructed distribution matches the true one, which requires focussing on small systems whose distribution is known at any time, we should emphasize that the ME procedure may also be most usefully employed when the true probability density is unknown. In such a case, it is implicitly assumed that the observables providing the constraints are easier to determine with accuracy than the distribution itself, thereby providing a way to reconstruct this distribution. It is nevertheless difficult in this case to assess the accuracy of the reconstruction resulting from the ME procedure.

In this paper we shall deal with the case where the constraints are provided by marginals instead of averages. Fortunately, an appropriate use of delta functions allows generalizing the previous results in a straightforward way. Taking for illustration the case $N=4$ (\textit{i.e.} $\mathbf x = (w,x,y,z)$), and assuming the tri-variate marginal $p_{123}(a,b,c)$ is known, putting $f(\mathbf x) = \delta(w,a)\delta(x,b)\delta(y,c)$ allows writing

\begin{align}
\langle f \rangle & = \sum_{\mathbf x} f(\mathbf x) p(\mathbf x) \notag \\
& = \sum_{w,x,y} \delta(w,a)\delta(x,b)\delta(y,c) \sum_z p(\mathbf x) \notag \\
& = \sum_{w,x,y} \delta(w,a)\delta(x,b)\delta(y,c) p_{123}(w,x,y)\notag \\
& = p_{123} (a, b, c) .
\end{align}
Applying the result above to all possible values of the arguments then yields

\begin{equation}
p(\mathbf x) = \frac{1}{Z} \exp \left( \sum_{a,b,c} \lambda(a,b,c) f(\mathbf x) \right) = \frac{1}{Z} \exp \left( \lambda(w,x,y) \right) ,
\end{equation}
where $\lambda$ now denotes a well-chosen \textit{function}. This generalizes to any number of marginals in a straightforward way; for instance, if besides $p_{123}$ the marginals $p_{124}$ and $p_{34}$ are given we get

\begin{equation}
p(\mathbf x) = \frac{1}{Z} \exp \left( \lambda_1 (w,x,y) + \lambda_2 (w,x,z) + \lambda_3 (y,z) \right)
\end{equation}
for some functions $\lambda_1, \lambda_2, \lambda_3$. Sadly, this elegant result is actually of little use since the determination of these $\lambda$'s is a difficult problem. An important exception is the simple case in which constrained marginals are univariate. Then $\lambda(w)=\ln p_1 (w)$ (and $\lambda(x)=\ln p_2 (x)$, \textit{etc.}) obviously satisfies the requirements, so that \textit{the ME distribution compatible with given univariate marginals is the factorized distribution}. In all other cases, we have to resort to the so-called \textit{iterative scaling algorithm} which allows numerical calculations.

Brown \cite{Brown1959} was among the first to describe this algorithm, the principle of which is, starting from some initial distribution, to consider all possible $n$-tuples of variables in sequence, each time adjusting the corresponding marginal. If we denote by $p^{(k)}$ the distribution obtained after $k$ such adjustments, $S_k$ the subset considered at the $k$-th step and $p_{S_k}$ the marginal distribution of this set, then the procedure is defined by

\begin{equation}
p^{(k)} := p^{(k-1)} \frac{p_{S_k}}{p^{(k-1)}_{S_k}} .
\end{equation}
The order in which the $n$-tuples are examined does not alter the distribution we converge to. From our experience, it seems that going twice through each $n$-tuple is enough to reach a satisfying solution. See \cite{Brown1959} for proofs of convergence. This scheme is demanding due to the fact that the number of $n$-tuples in a set grows factorially.

\section{III. Decomposing multi-information}

When looking at the ME reconstruction based on the knowledge of, say, tri-variate marginals, it is not easy \textit{a priori} to know if it is accurate because it takes into account strong triplet-wise interactions between variables, or if the same could have been achieved by the reconstruction based on bi-variate marginals only. We need a way to disentangle different orders of interaction, as well as a tool to compare distributions.

Both are provided by the \textit{Kullback-Leibler (KL) divergence} between two distributions $p$ and $q$ (living on the same support), which is defined as

\begin{equation}
D(p,q) := \sum_{\mathbf x} p(\mathbf x) \ln \frac{p (\mathbf x)}{q (\mathbf x)} .
\end{equation}
This quantity provides a pseudo-distance on the space of probability distributions \cite{Cover2006}.

The idea now is to use the KL divergence to compare the distribution we want to approximate with the ME distribution based on marginals of a certain order $k$ by computing $D(p,p_{ME}^{(k)})$, where $p_{ME}^{(k)}$ denotes the ME reconstruction based on $k$-marginals. It is intuitively clear that the larger the subsets considered, the more accurate the resulting ME approximation will be (indeed smaller-order marginals may be recovered from larger-order ones), whence the inequality $D(p,p_{ME}^{(k-1)}) \geq D(p,p_{ME}^{(k)})$. The difference may therefore be interpreted as the gain in accuracy when basing our guess on subsets of size $k$ instead of $k-1$, and therefore quantifies specifically the role played by interactions of order $k$. We define accordingly

\begin{equation}
C_k := D(p,p_{ME}^{(k-1)}) - D(p,p_{ME}^{(k)}) \geq 0 .
\end{equation}
This quantity is sometimes referred to as the \textit{connected multi-information of order $k$} \cite{Schneidman2003}, but in our opinion this name is unfortunate (where does connectedness enter into the play ?) and will not be used here.

Performing the telescopic sum of all these coefficients gives

\begin{equation}
\sum_{k=2}^N C_k = D(p,p_{ME}^{(1)}) - D(p,p_{ME}^{(N)}) .
\end{equation}
Since $p_{ME}^{(N)}$ is trivially $p$ itself (whence $D(p,p_{ME}^{(N)}) = 0$), and since, as mentioned above, $p_{ME}^{(1)} = \prod_{i=1}^N p(x_i)$, we get finally

\begin{align}
\sum_{k=2}^N C_k & = D \left( p(\mathbf x), \prod_{i=1}^N p(x_i) \right) \notag \\
& =\sum_{\mathbf x} p(\mathbf x) \ln \frac{p(\mathbf x)}{\prod_{i=1}^N p(x_i)} := M.
\end{align}
The quantity $M$ is known as the \textit{multi-information} \cite{Watanabe1960} and quantifies the total amount of interdependence inside a set of variables. (As an aside, note that the sum could be started from $k=1$, in which case it would add up to the KL distance between $p$ and the uniform distribution. The total amount of interdependence could indeed be defined that way, but it is usual to say that this quantity is given by the multi-information as defined above. Moreover, the standard convention attributes null interdependence to a set of \textit{independent} variables which would not be the case otherwise, where only \textit{uniform} distributions would be said to have no interdependence. This alternative way to define things could nevertheless be considered occasionally.) This formula shows, as could have been expected, that the total interdependence in a system is built by addition of pair-wise, triplet-wise, and so on, interactions.

An important question is whether all subsets of a given order should be treated on the same footing when there exists a notion of distance between variables, as would be the case for instance if the system were put on a graph and each variable assigned to a node (more generally, such a distance \textit{has} to exist as soon as a focus is put on ``multi-scaleness''). Considering for instance the case of pairs, co-dependence between variables remote from each other will intuitively be much smaller than between two neighbouring variables, so that it seems acceptable to discard pairs constituted by distant elements. On another side, these loose pairs are by far more numerous than tightly-bound ones, so that though their contribution is weak envisaged individually, it cannot be neglected anymore when considered globally. But this very fact that pairs (more generally $n$-tuples) are so many implies that considering them all becomes numerically difficult (factorial growth). Since in the experiments to follow we focus on systems which display such a notion of distance, provided by the topology of the graph on which we put our variables, our viewpoint will be to consider as admissible $n$-tuples only those consisting of connected variables (\textit{i.e.} $n$-tuples of which restricted graph is connected). A few checks (see below) tend to suggest than our conclusions are not drastically altered in the more general case where all subsets are retained. We have to admit that this simplification is questionable and intend to address more specifically this issue elsewhere.

\section{IV. Elementary cellular automata and Langton's parameter}

One-dimensional ECA have been very thoroughly investigated (see \cite{Chopard1998} for an introduction). Wolfram \cite{Wolfram1983} noted that these may be classified in four different classes: class I regroups ECA converging to some homogeneous pattern; class II displays an inhomogeneous stable pattern or periodic behaviour; class III displays completely chaotic behaviour; finally class IV regroups automata which exhibit slowly building up and decaying sub-patterns. It is believed that ECA belonging to class IV are the closest to our intuitive conception of complexity. While this classification is very widely used, many alternatives have been proposed (see \cite{Martinez2013} and references therein). One of these alternatives is to consider \textit{Langton's $\lambda$ parameter}, which provides a parametrization of the space of cellular automata. In the case of elementary one-dimensional automata, it is computed very easily as the percentage of configurations in the lookup table giving rise to a living cell (\textit{i.e.} a cell taking value 1), but this parameter may be generalized to any kind of automaton \cite{Langton1990}. A \textit{leitmotiv} of Langton's reflexion was to suggest that, while simple (classes I and II) automata correspond to small values of $\lambda$ and chaotic ones (class III) to values close to $\lambda = 1/2$, complex CA should emerge somewhere in between, at what has been popularized as the \textit{edge of chaos}.

The idea of quantifying the role played by successive orders of interaction in the informational content of cellular automata has already been adressed by Lindgren and Nordahl \cite{Lindgren1987, Lindgren1988}. However, these authors do not resort to ME methods but use instead a simpler decomposition of the entropy rate, which makes this tool restricted to one-dimensional topologies (this limitation of their work was actually an important motivation for undertaking the present study).

\section{V. Computational framework}

Before moving on to discuss our results, we should say a few words about our computational framework and address how the temporal evolution of the probability density is handled. Often this is done by means of Monte-Carlo methods, by evolving copies of the system and reconstructing the probabilities by sampling trajectories (following this approach see \cite{Quax2013} for a recent work on a closely related topic). Here we follow an alternative approach which is to determine the time evolution of the probability density exactly. Then we may, so to speak, follow simultaneously all trajectories down to the least probable ones. This amounts to a description in terms of Markov chains, where the knowledge of history allows to predict towards which states the system could evolve. We will restrict ourselves here to the case where the knowledge of a \textit{finite} history is sufficient to predict possible futures. By suitably extending the state space, actually all such processes may be recast in the form of \textit{memoryless Markov chains} (or simply \textit{Markov chains}), by what we mean that the probabilities of the forthcoming states may be predicted knowing the current state of the process only. In the case of ECA considered here, the Markov process is memoryless by construction.

While this approach seems to outperform sampling in terms of accuracy, it actually suffers from its numerical cost when the system's size increases. Assuming for instance we deal with a system constituted by $N$ variables taking binary values, we have in this case $2^N$ possible configurations, while assuming the system is driven by a dynamics with a $k$-steps memory yields $2^{kN}$ possible histories to keep into consideration, which becomes soon untractable even for small values of $N$ and $k$.

Nonetheless this formalism has some advantages which justify its adoption in this paper. In particular it allows a more straightforward transition from numerical exploration to theoretical investigation. Still more importantly, as we already mentioned, this approach does not require to select (arbitrarily) an initial configuration, but handles them all as long as they are not explicitly assigned probability zero from the beginning. This will turn out to be a crucial feature.

\section{VI. Results}

We first computed the time evolution of $C_k$ coefficients for ECA of size $N=10$ put on a periodic string-like topology, as well as the behaviour of the multi-information. Two instances are presented below. Figure \ref{r110a} shows the result for rule 110, which is a most typical instance of a CA displaying W-complex behaviour. Note that all coefficients except $C_9$ and $C_{10}$ provide a significant contribution to the multi-information (note that the coefficients are displayed cumulatively; see caption). Remark that after a transient phase of around ten steps, the system enters a periodic regime. While the multi-information is almost constant in this regime, the respective contributions present a much stronger variability. $C_4$, for instance, alternatively reaches significant value and then decays close to zero. On the contrary, the contribution of $C_8$ is nearly constant in the stationary regime after reaching a peak during the transient phase.

\begin{figure}
\includegraphics[scale=0.4]{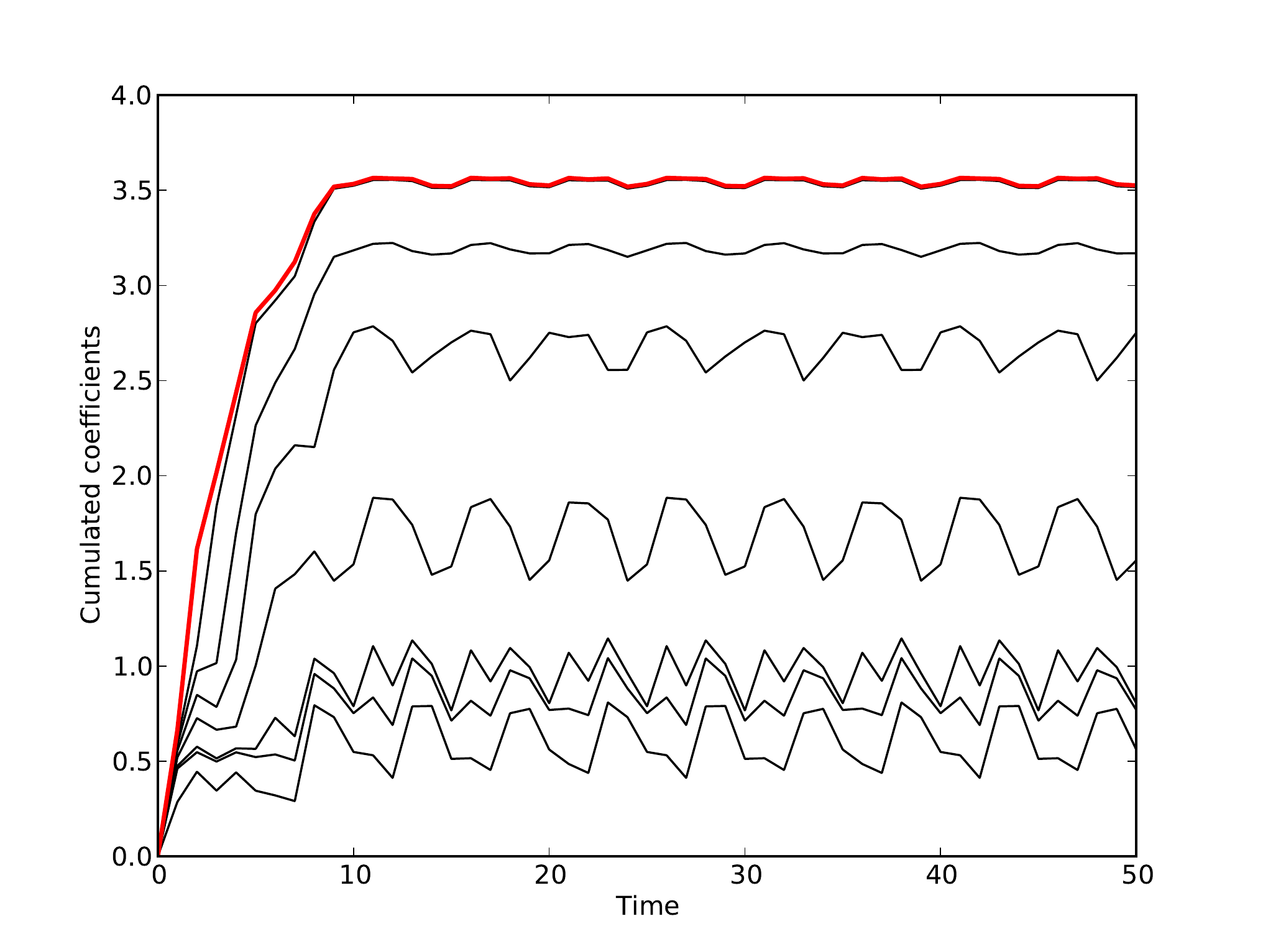}
\caption{\small Time evolution of $C_k$ coefficients in rule 110 (adjacent subsets). Coefficients are displayed cumulatively, \textit{i.e.} we show successively (from bottom to top) $C_2$, $C_2+C_3$, \textit{etc.} The sum converges to $M$ (red curve).}
\label{r110a}
\end{figure}

\begin{figure}
\includegraphics[scale=0.4]{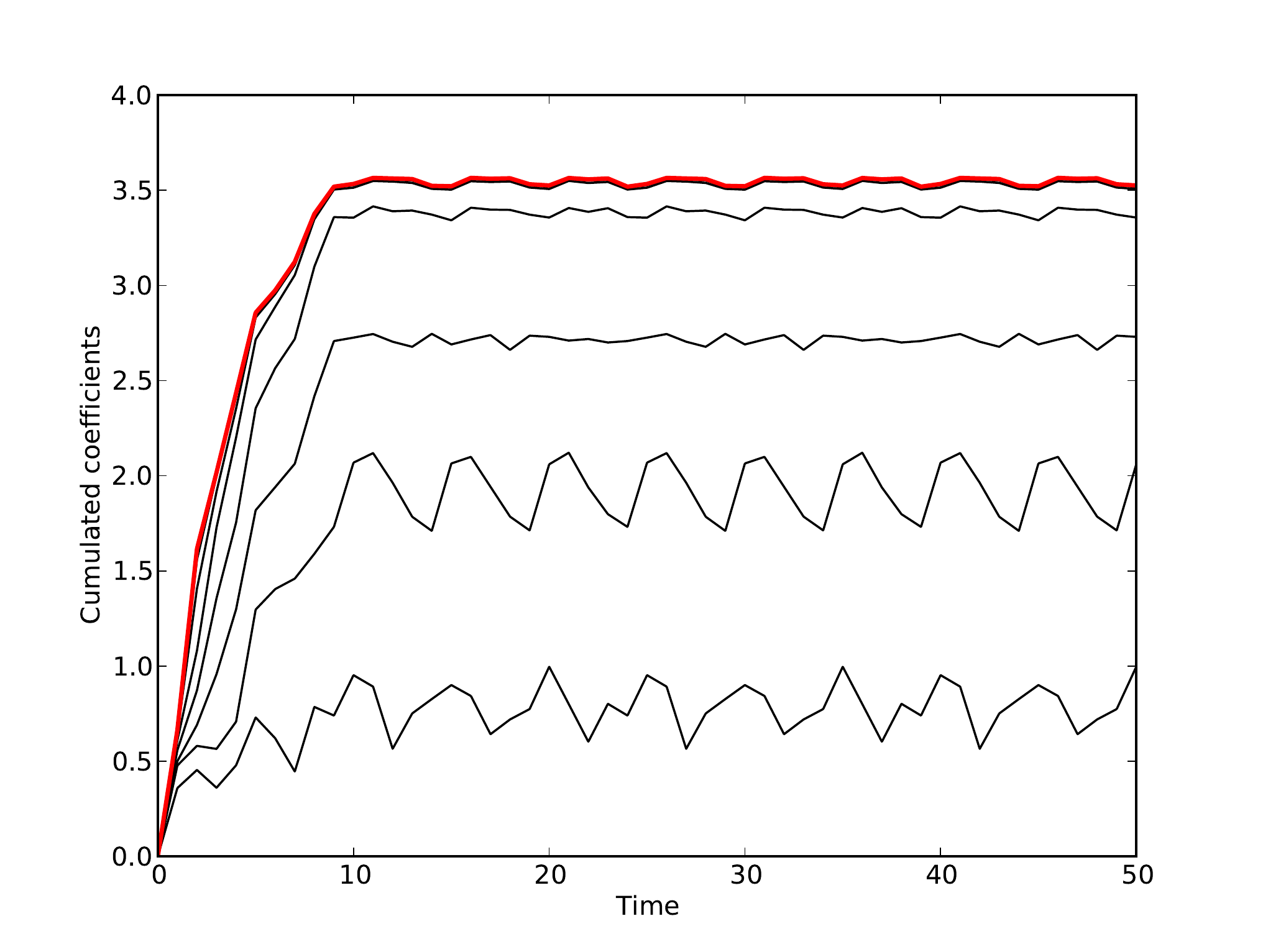}
\caption{\small Time evolution of $C_k$ coefficients in rule 110 (all subsets)}
\label{r110t}
\end{figure}

For comparison we show in figure \ref{r110t} the corresponding picture when all subsets of a given order, instead of adjacent ones only, are taken into account. The main difference is that $C_7$ and $C_8$ play no significant role, but on the whole the behaviour of the remaining coefficients is not qualitatively altered. Note however how our decision to rule out non-adjacent subsets introduced spurious high-order dependences.

\begin{figure}
\includegraphics[scale=0.4]{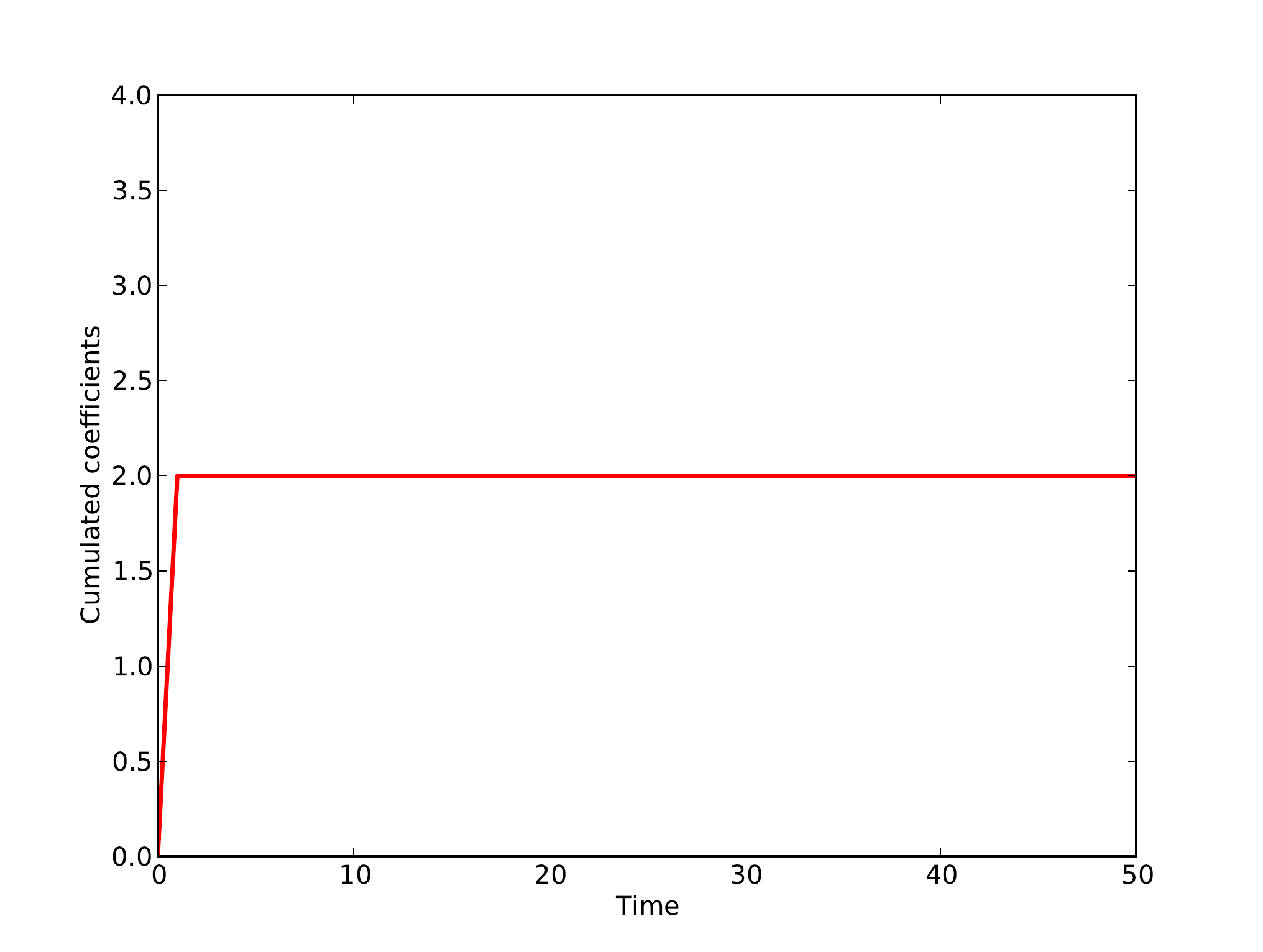}
\caption{\small Time evolution of $C_k$ coefficients (adjacent subsets) in rule 90. Only $C_9$ contributes.}
\label{r90a}
\end{figure}

This behaviour is in sharp constrast with the one shown in figure \ref{r90a}, which displays the same plot for rule 90. This rule is characterized by the fact that the sole contribution to multi-information is provided by the coefficient $C_9$, all other orders of interaction vanishing (admittedly this could hardly be guessed from the plot alone). Very interestingly, rule 90 is nothing but the dynamics obtained by applying the XOR operator on the two neighbours of a variable and assigning the result to the variable. This highlights the fact that the notion of ``order of interaction'' as employed in the current context does not quite overlap what we could expect from, for instance, classical kinetic theory. There, ``interaction of order $n$'' would be understood in terms of the functional form of the energy function (in the sense that the latter could not be decomposed as a sum of functions involving less than $n$ variables). The case of rule 90 would then correspond to interactions of order two, and ECA in general to interactions of order three. It would then be difficult to justify the appearance of higher orders of interaction. An interesting question is to find a dynamics such that all informational content is brought in by interactions of order $k$, \textit{i.e.} $C_k=M$ and $C_l=0$ $\forall l \neq k$.

\begin{figure}
\includegraphics[scale=0.4]{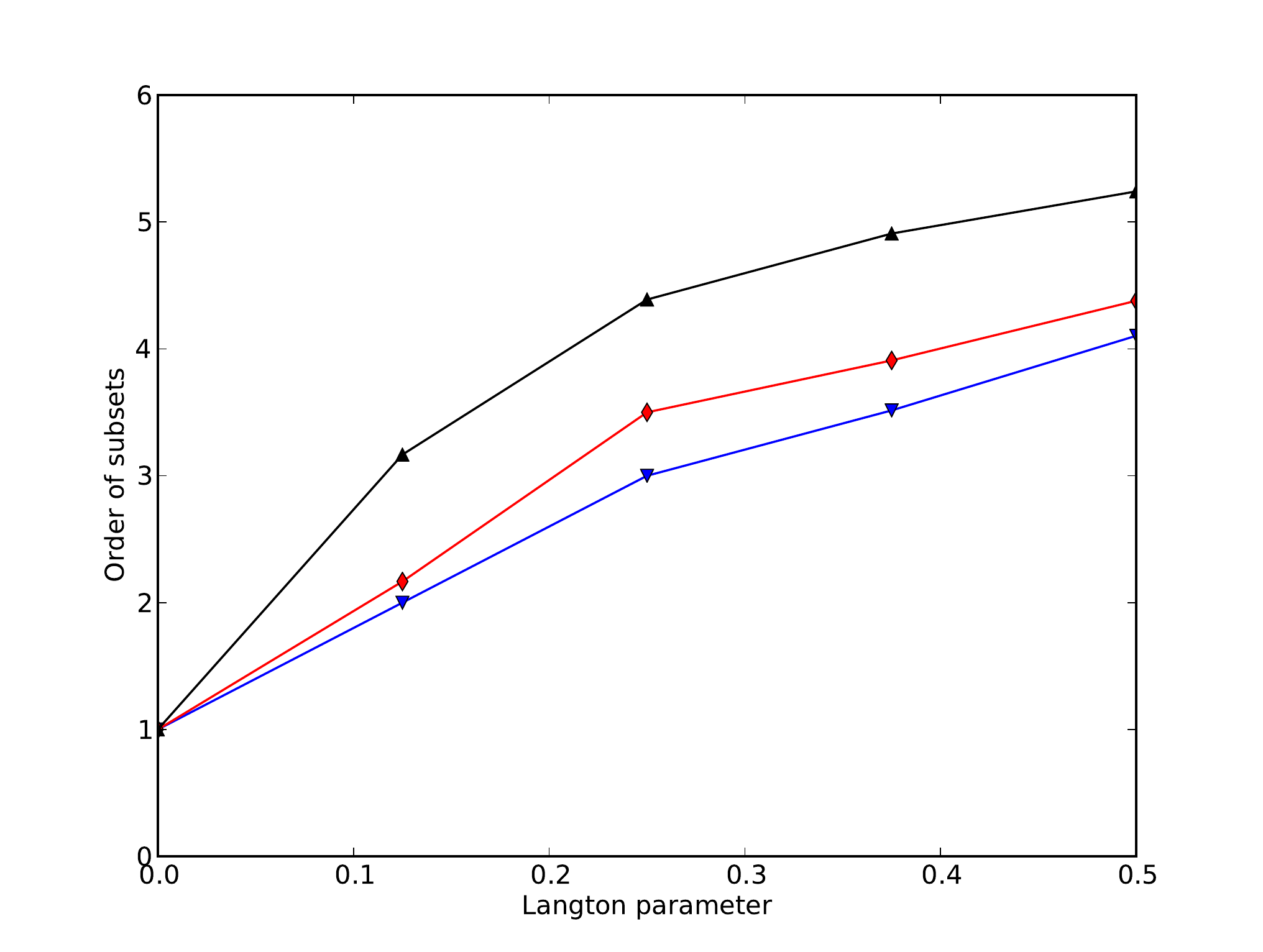}
\caption{\small Size of subsets required to reconstruct a ME distribution recovering 50\%, 70\% and 90\% of the multi-information respectively}
\label{langton10}
\end{figure}

We should actually better get rid of the transient phase, details of which depend on the probabilities we initially assigned to each configuration. Moreover, in order to get more easily displayable results, we shall discard temporal variations of $C_k$ coefficients. From now on, coefficients will therefore be averaged (when necessary) over the stationary phase, so that all forthcoming statements about $C_k$ coefficients will be statements about the average value of these coefficients in the stationary regime. Figure \ref{langton10} displays, for each value of the Langton parameter $\lambda$, up to what size $\langle \langle \Sigma \rangle_T \rangle_{\lambda}$ the subsets should be considered in order to recover, respectively, 50\%, 70\% and 90\% of the multi-information in the stationary regime (while heavy, this notation has the merit to make clear that this is the size obtained from coefficients averaged over time \textit{and} over all rules having the same $\lambda$). We only consider the range $\lambda \in [0,1/2]$ since any ECA with $\lambda > 1/2$ is mirrored by another one which is equivalent but has $\lambda' = 1 - \lambda$. We observe that $\langle \langle \Sigma \rangle_T \rangle_{\lambda}$ increases monotonically with $\lambda$, which means that A-complexity grows with $\lambda$, and this statement holds whatever the percentage of multi-information targeted.

\begin{figure}
\includegraphics[scale=0.4]{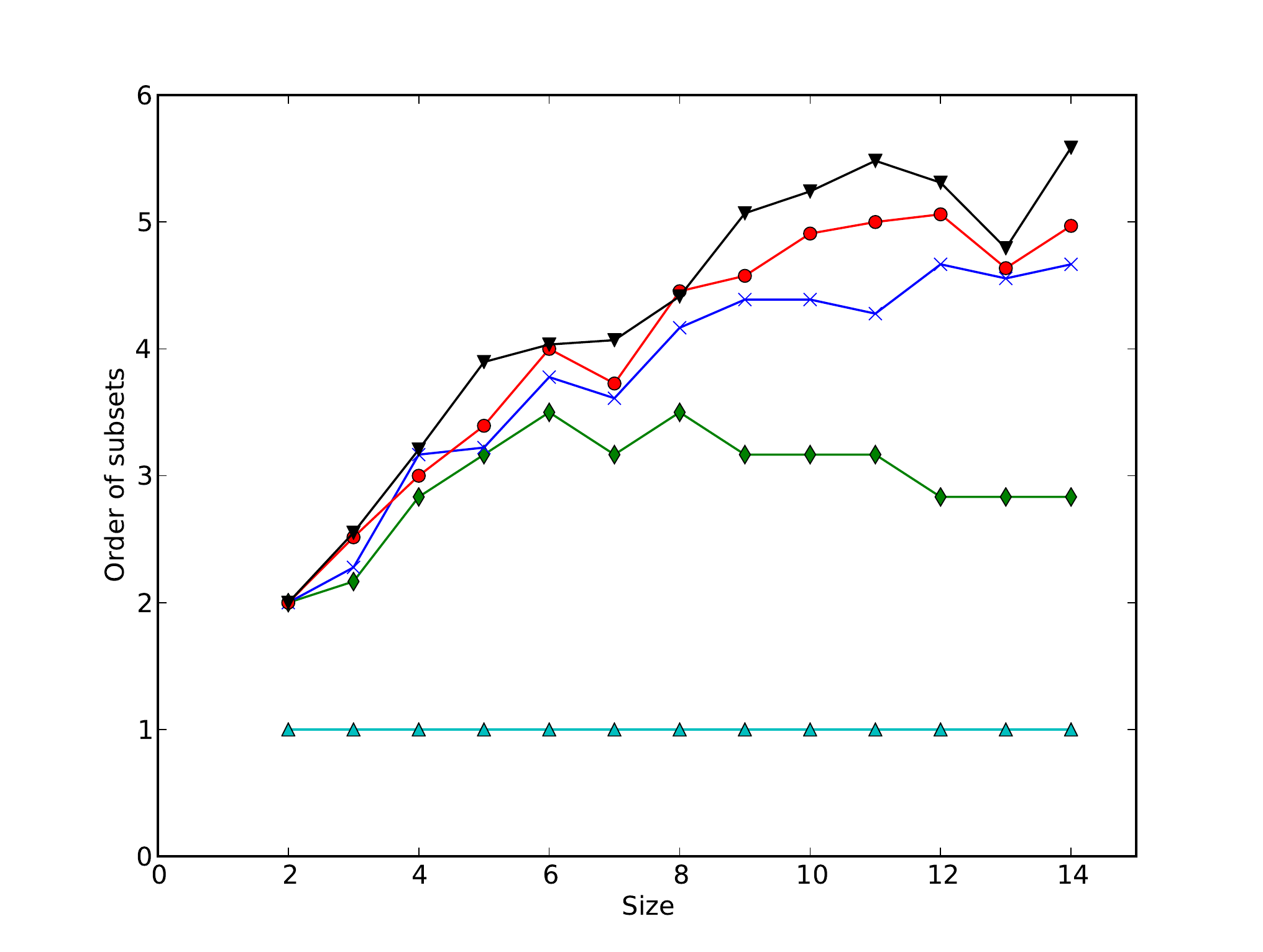}
\caption{\small Order of subsets required to recover 90\% of the multi-information as a function of $N$, for each value of $\lambda$ ($\lambda=0$: cyan; $\lambda=1/8$: green; $\lambda=1/4$: blue; $\lambda=3/8$: red; $\lambda=1/2$: black)}
\label{newavsizes88}
\end{figure}

We now turn to the dependence of $\langle \langle \Sigma \rangle_T \rangle_{\lambda}$ on the size of the ECA, for the case where a 90\% reconstruction of the multi-information is targeted. Results are displayed in figure \ref{newavsizes88}. The curve for $\lambda=0$  is special since it comprises only one rule (R0). For this rule considering individual elements -actually only one of them- is enough to characterize the system, so that in this case $\langle \langle \Sigma \rangle_T \rangle_{\lambda=0}=1$ whatever the size of the system. For other values of $\lambda$, $\langle \langle \Sigma \rangle_T \rangle_{\lambda}$ grows with $N$, in a way which is investigated below.

The fact that $\langle \langle \Sigma \rangle_T \rangle_{\lambda}$ increases monotonically with $\lambda$ seems to be common to all sizes, except for $N=4$ and $N=8$ by a small amount. Though it seems difficult to single out one specific cause for this inversion, the forthcoming discussion should make the issue clearer.

Figure \ref{newavsizes88} says nothing about possible heterogeneity inside the set of rules having the same $\lambda$. It will turn out that this intra-$\lambda$ variability is indeed extremely important, so that a discussion in terms of $\lambda$ is actually rather irrelevant. We will therefore now discuss the question in detail by examining rules for themselves, without trying to tie links to the $\lambda$-parametrization.

\begin{figure*}
        \centering
        \begin{subfigure}[b]{0.3\textwidth}
                \includegraphics[scale=0.25]{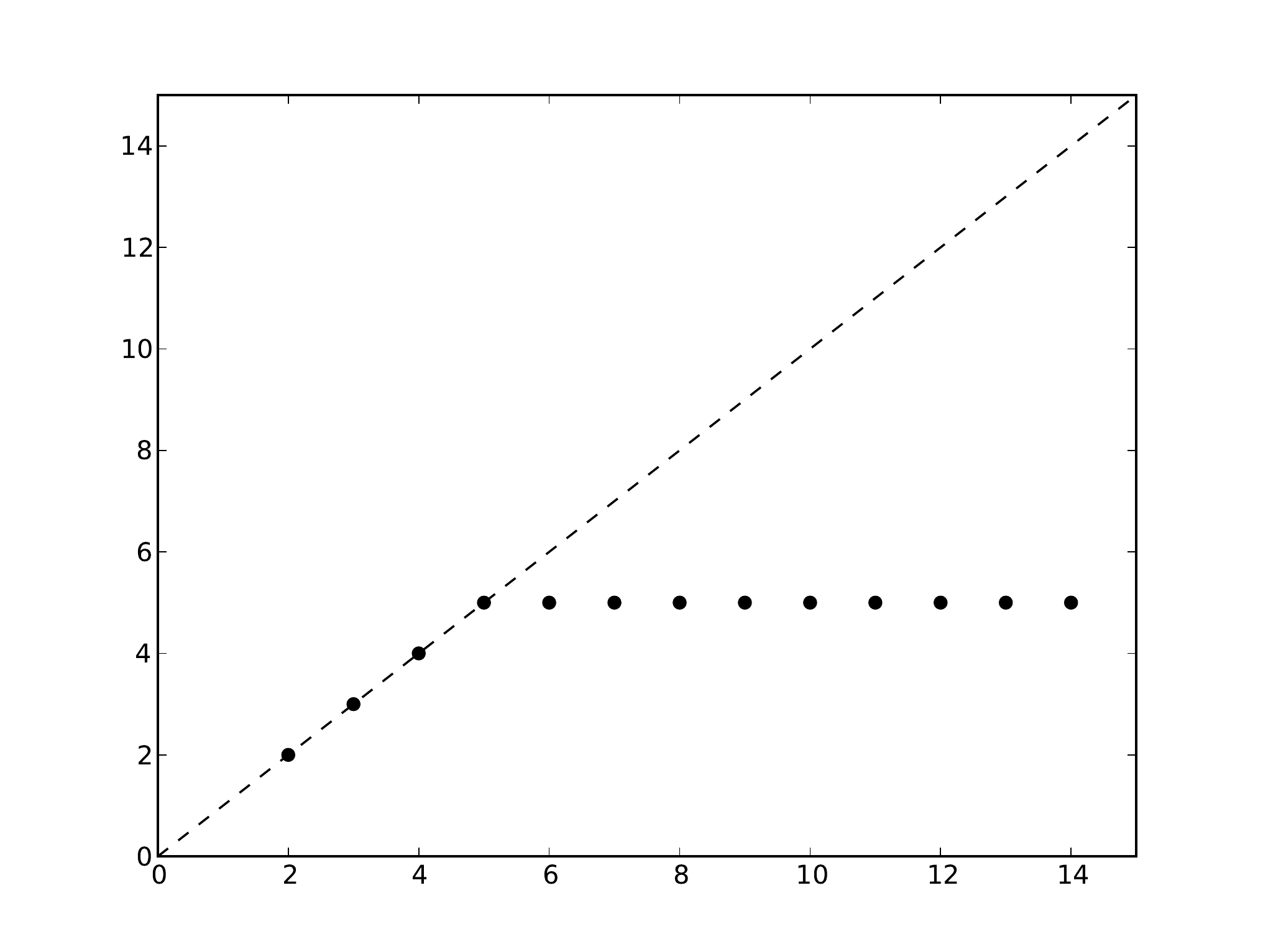}
                \caption{R14}
        \end{subfigure}
        \quad
        \begin{subfigure}[b]{0.3\textwidth}
                \includegraphics[scale=0.25]{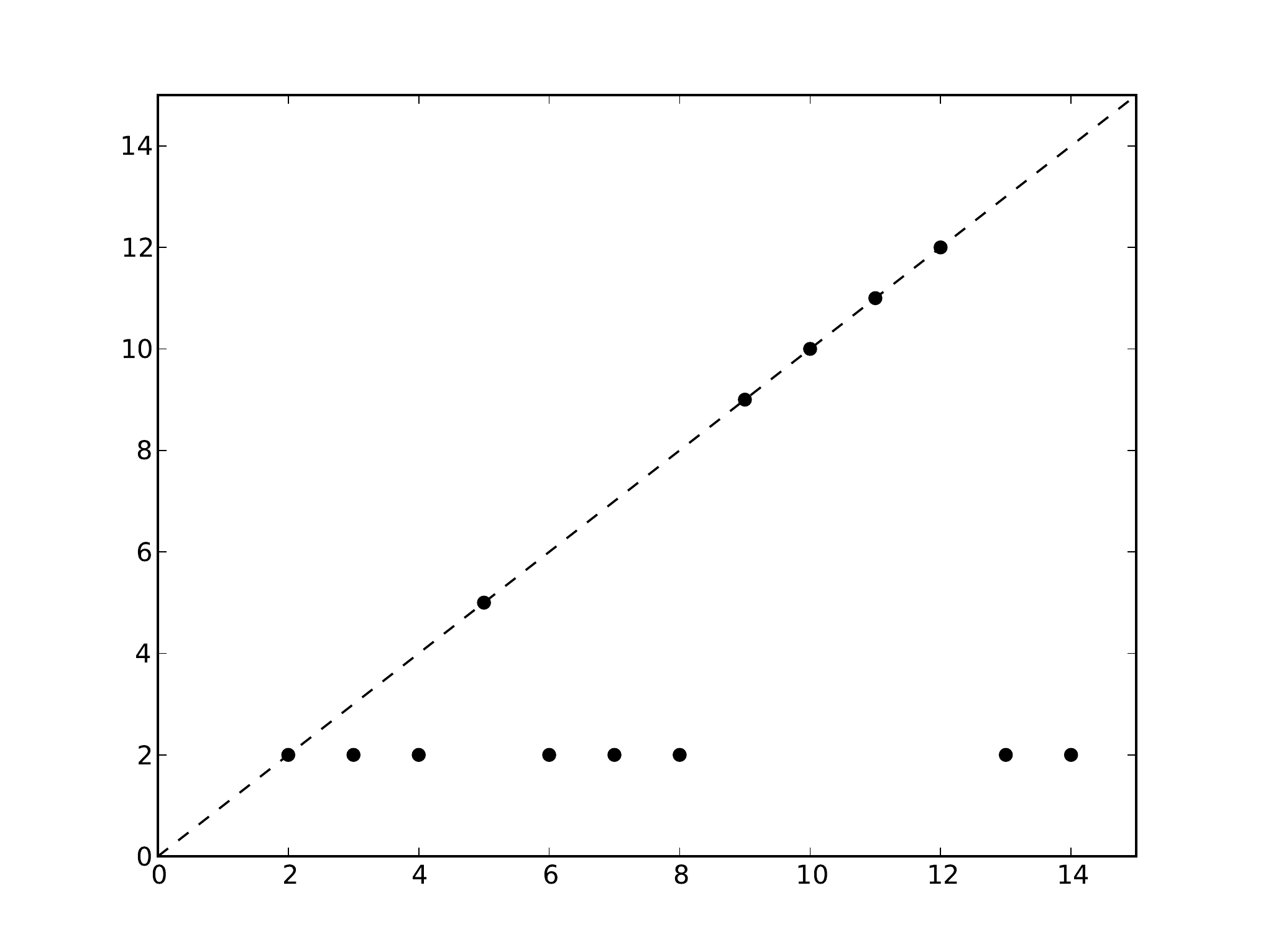}
                \caption{R15}
        \end{subfigure}
	\quad
        \begin{subfigure}[b]{0.3\textwidth}
                \includegraphics[scale=0.25]{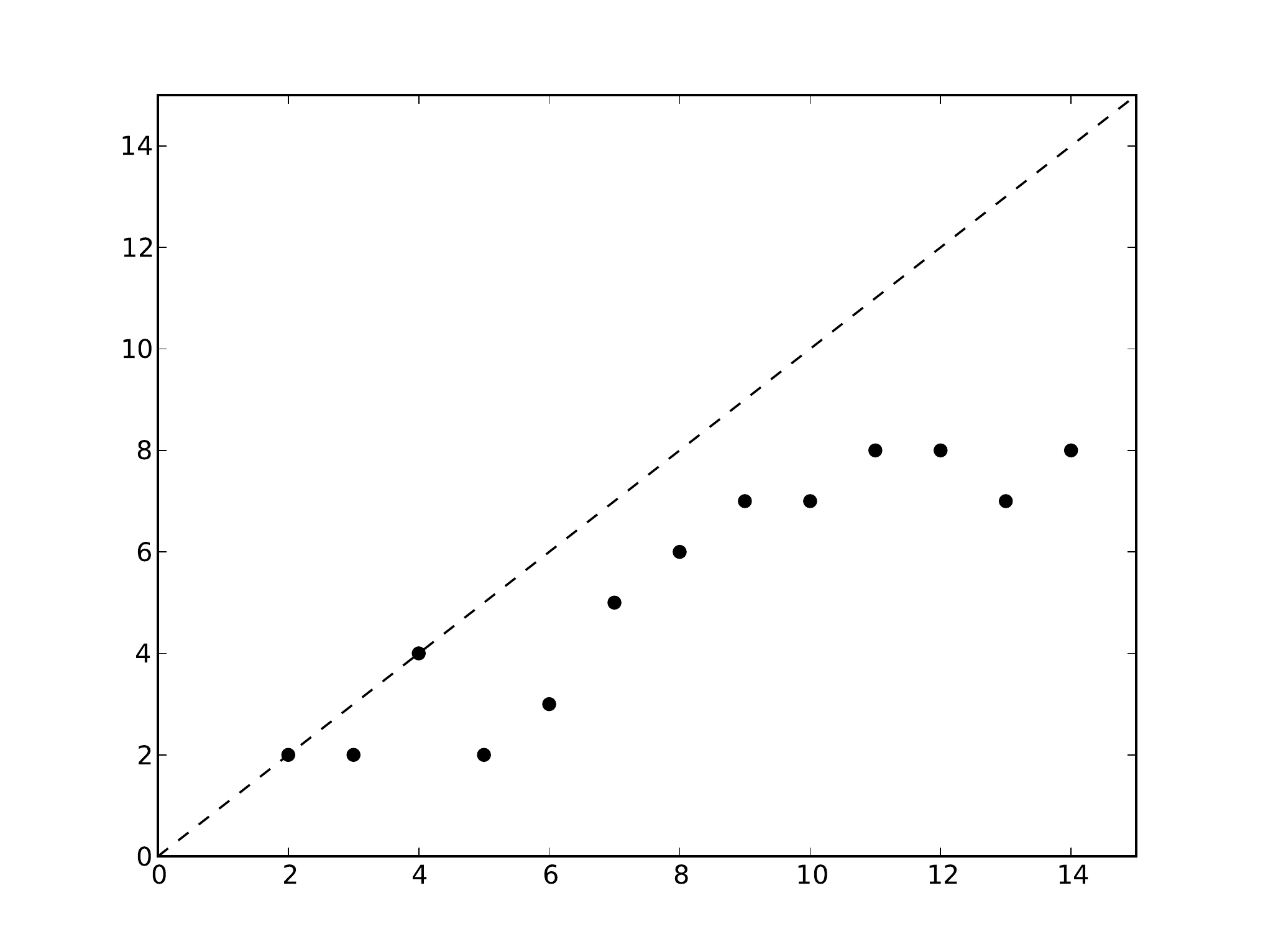}
                \caption{R22}
        \end{subfigure}

        \begin{subfigure}[b]{0.3\textwidth}
                \includegraphics[scale=0.25]{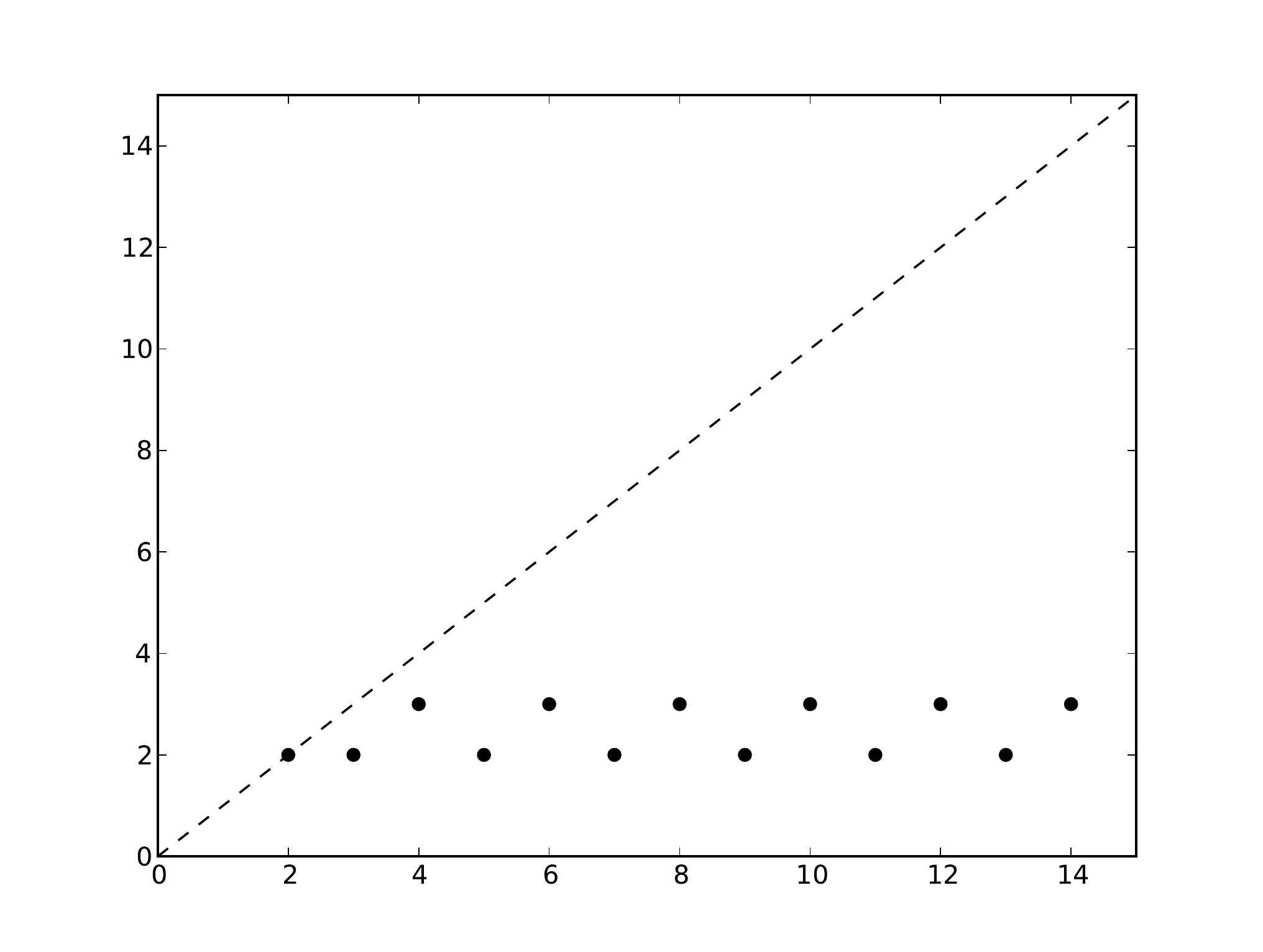}
                \caption{R32}
        \end{subfigure}
	\quad
	\begin{subfigure}[b]{0.3\textwidth}
                \includegraphics[scale=0.25]{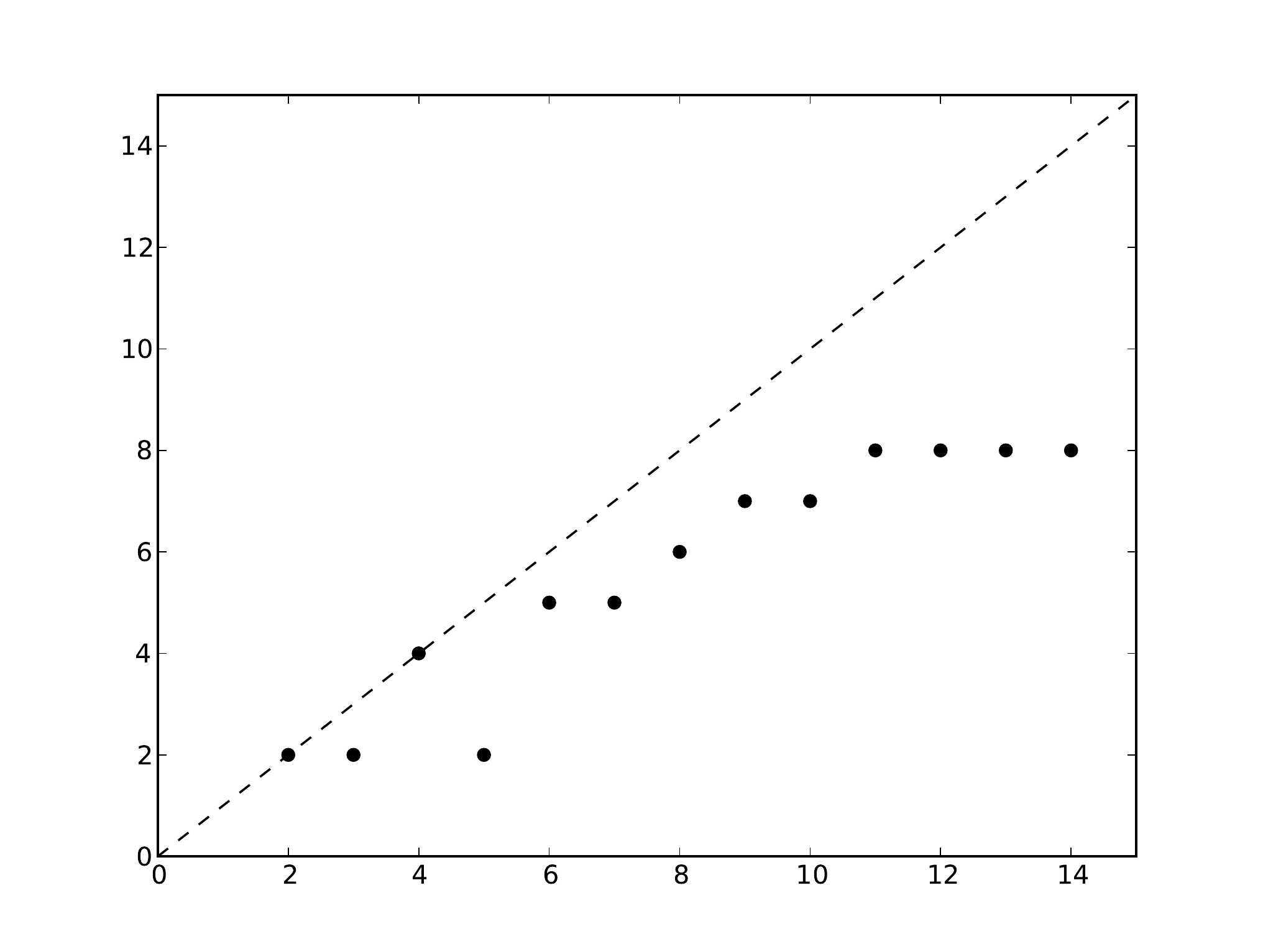}
                \caption{R54}
        \end{subfigure}
	\quad
        \begin{subfigure}[b]{0.3\textwidth}
                \includegraphics[scale=0.25]{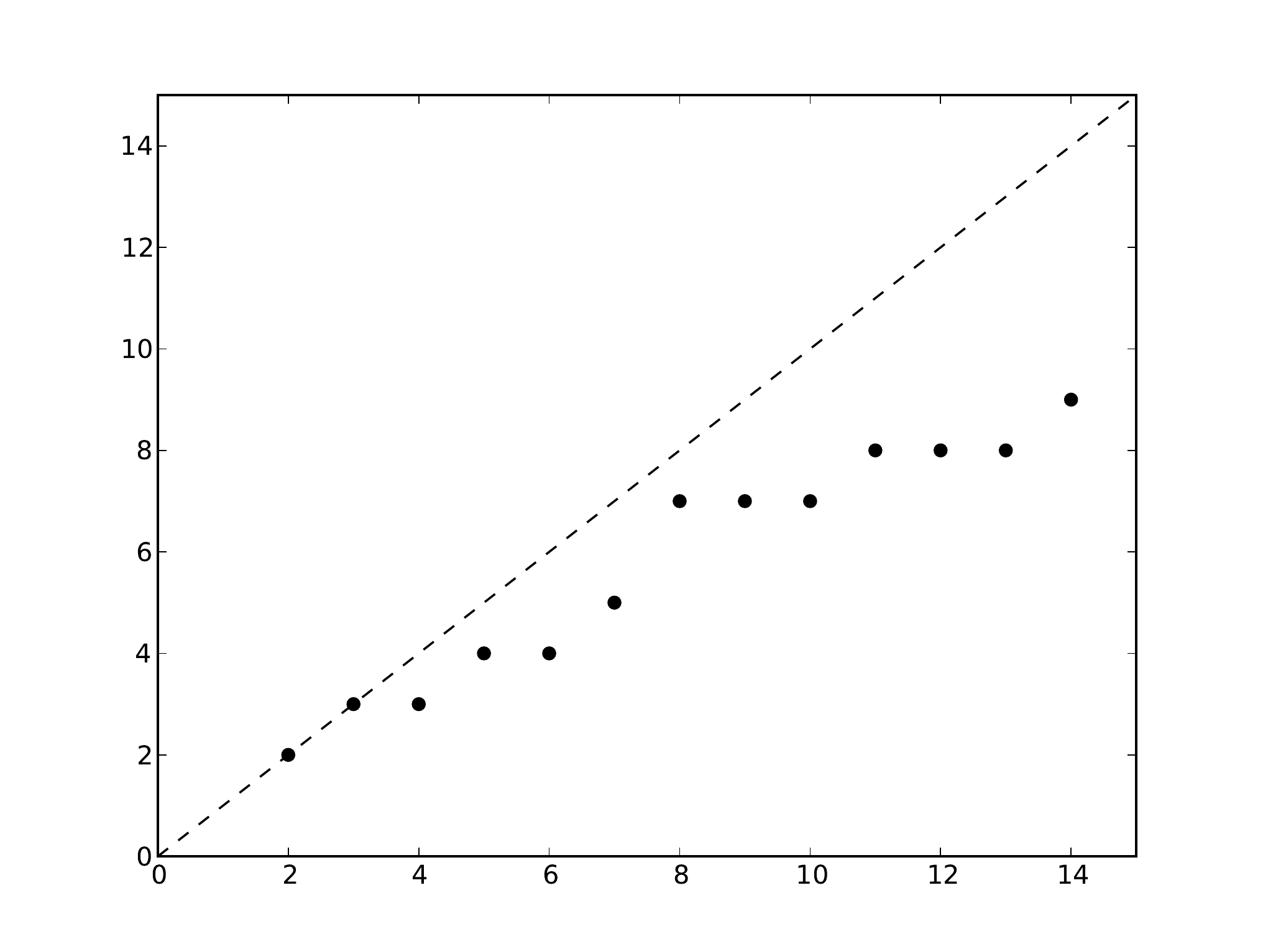}
                \caption{R73}
        \end{subfigure}

	\begin{subfigure}[b]{0.3\textwidth}
                \includegraphics[scale=0.25]{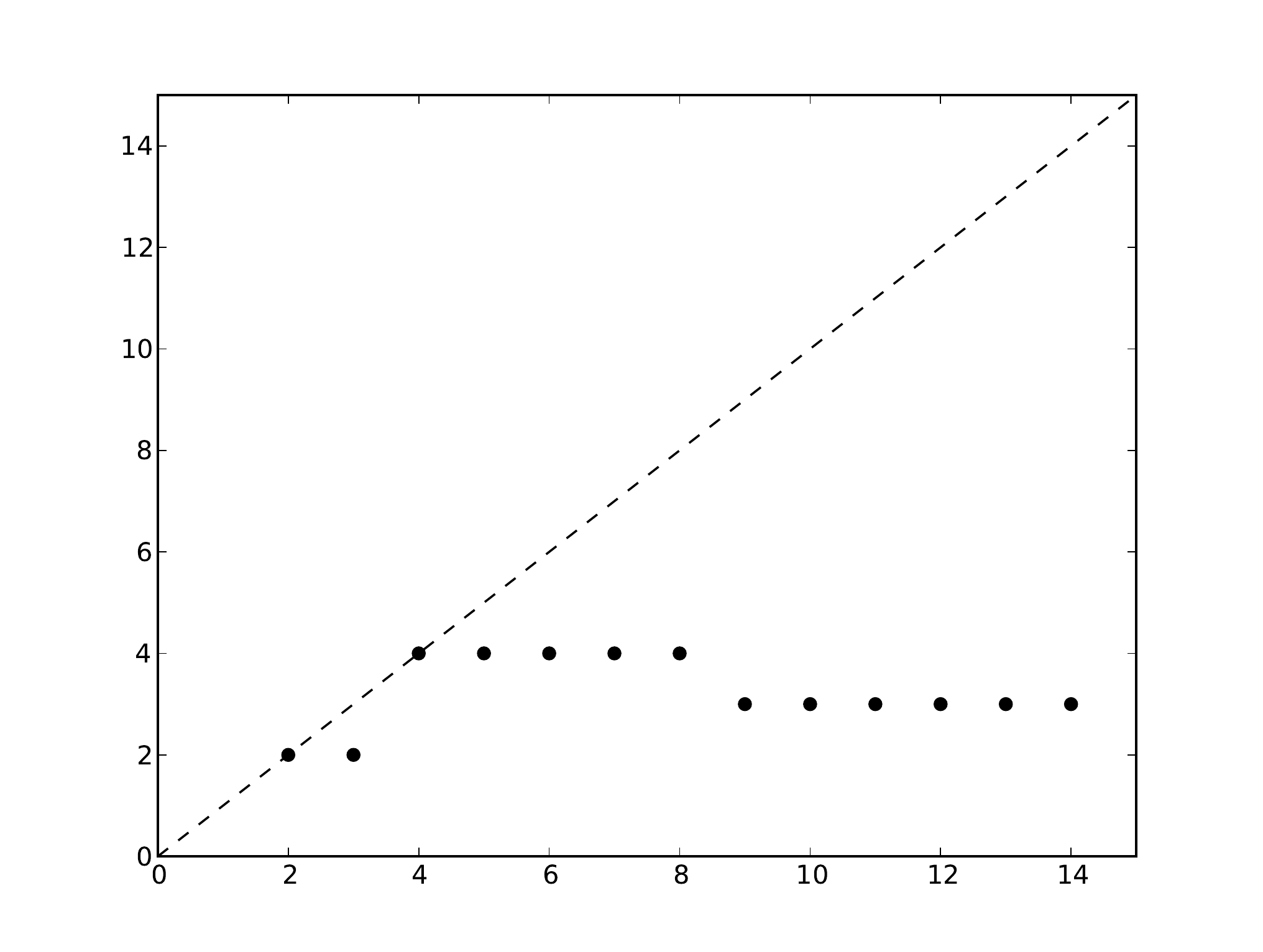}
                \caption{R77}
        \end{subfigure}
	\quad
        \begin{subfigure}[b]{0.3\textwidth}
                \includegraphics[scale=0.25]{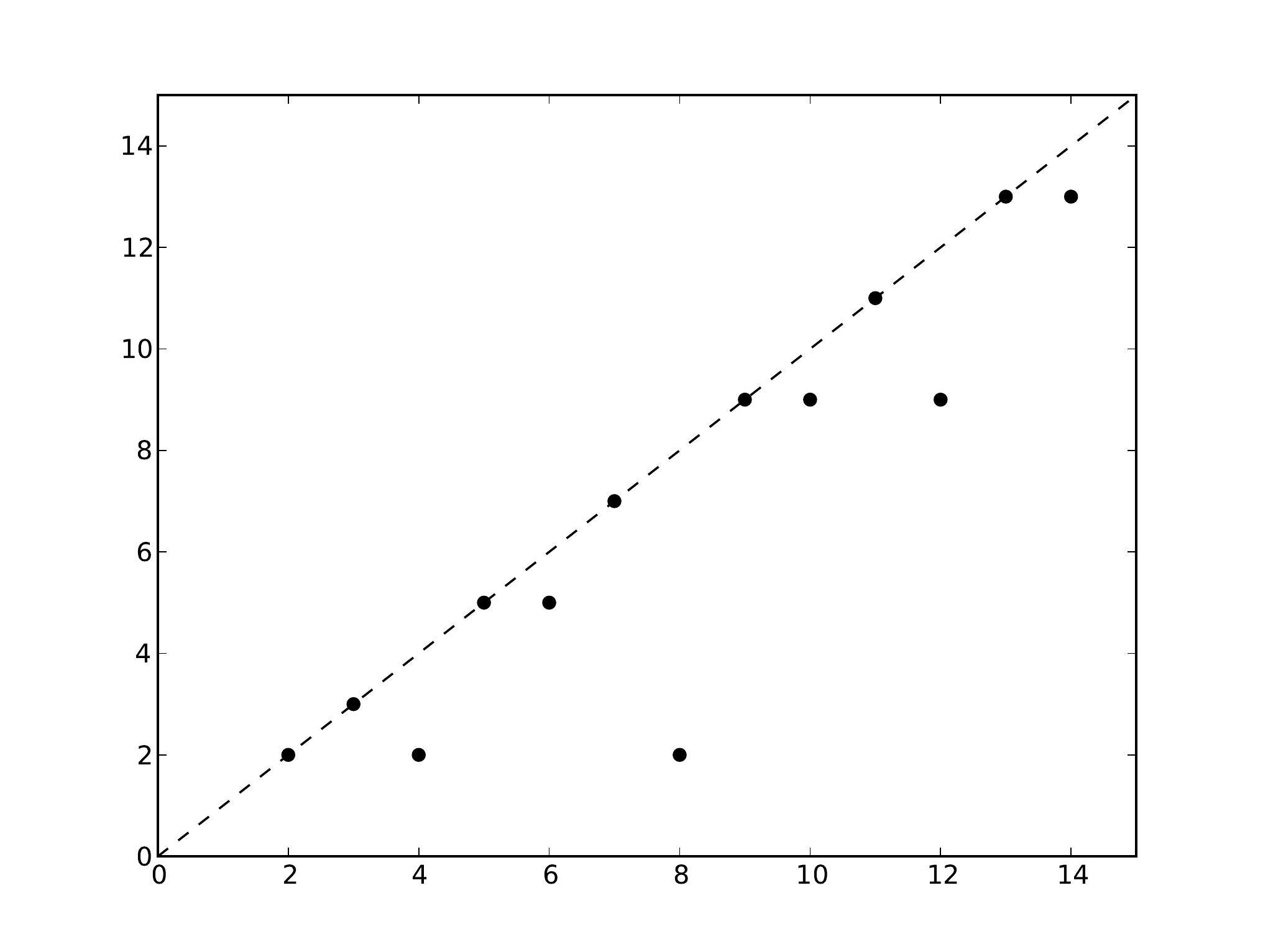}
                \caption{R90}
        \end{subfigure}
	\quad
	\begin{subfigure}[b]{0.3\textwidth}
                \includegraphics[scale=0.25]{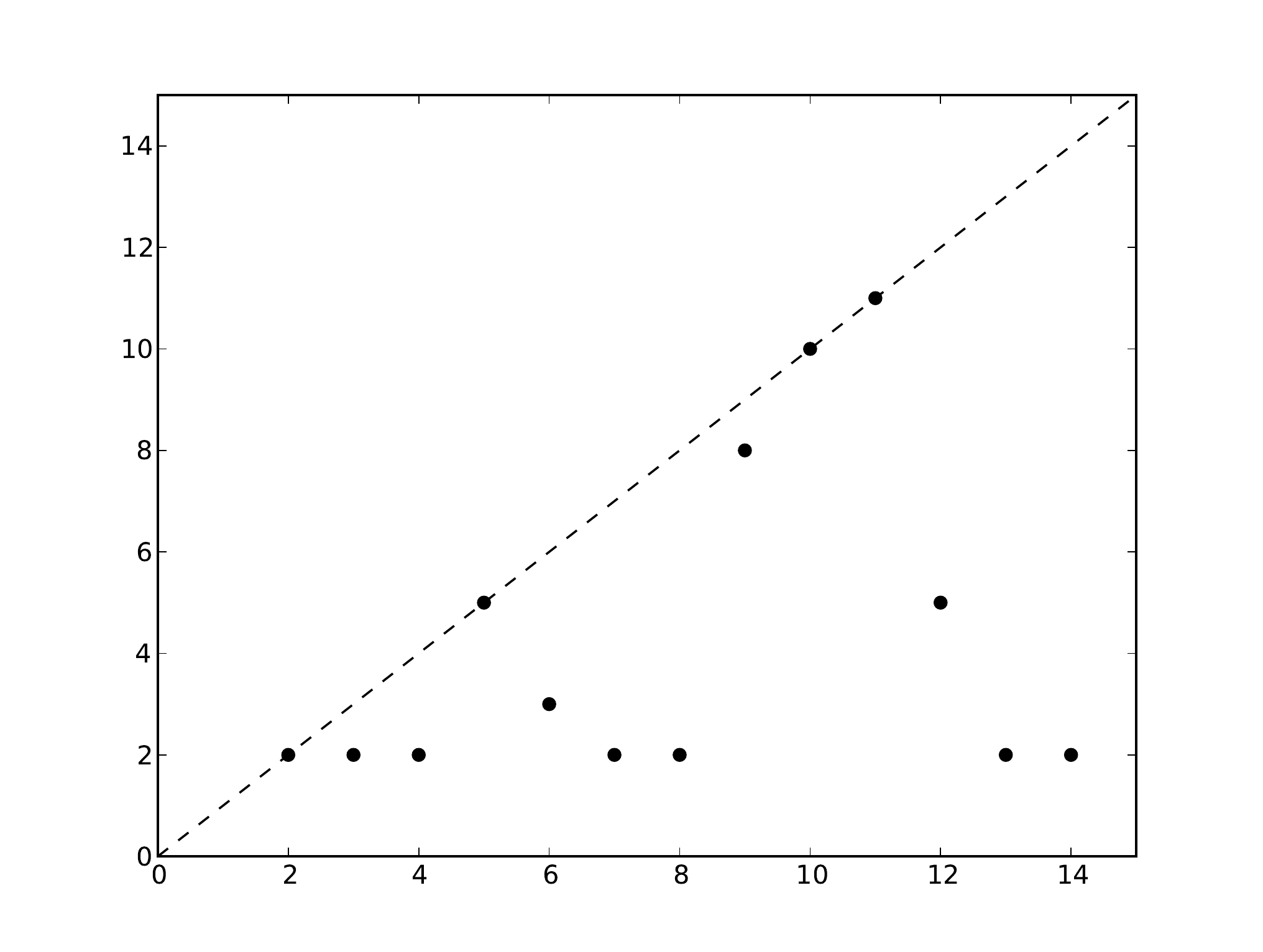}
                \caption{R105}
        \end{subfigure}

        \begin{subfigure}[b]{0.3\textwidth}
                \includegraphics[scale=0.25]{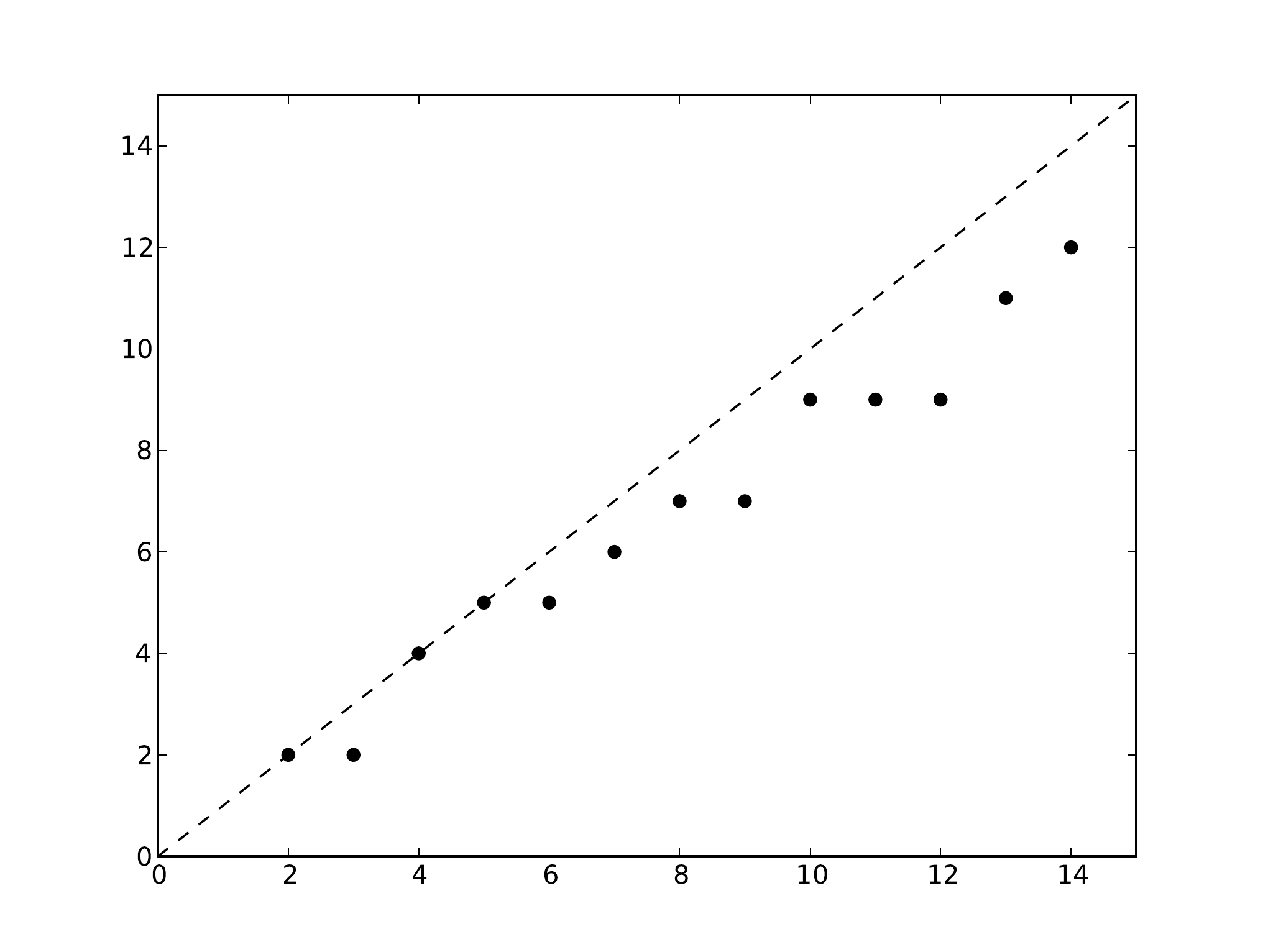}
                \caption{R106}
        \end{subfigure}
	\quad
	\begin{subfigure}[b]{0.3\textwidth}
                \includegraphics[scale=0.25]{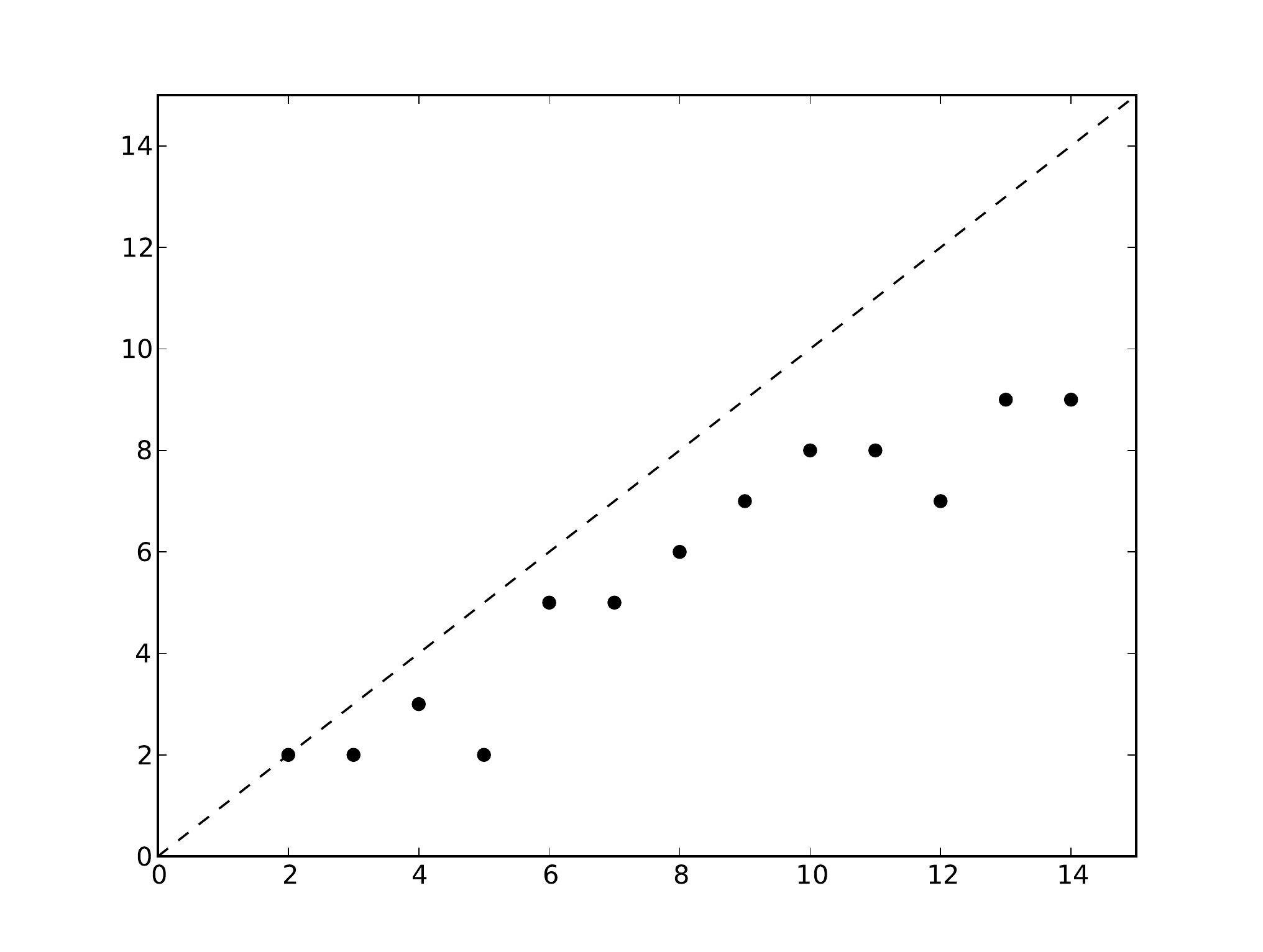}
                \caption{R110}
        \end{subfigure}
	\quad
        \begin{subfigure}[b]{0.3\textwidth}
                \includegraphics[scale=0.25]{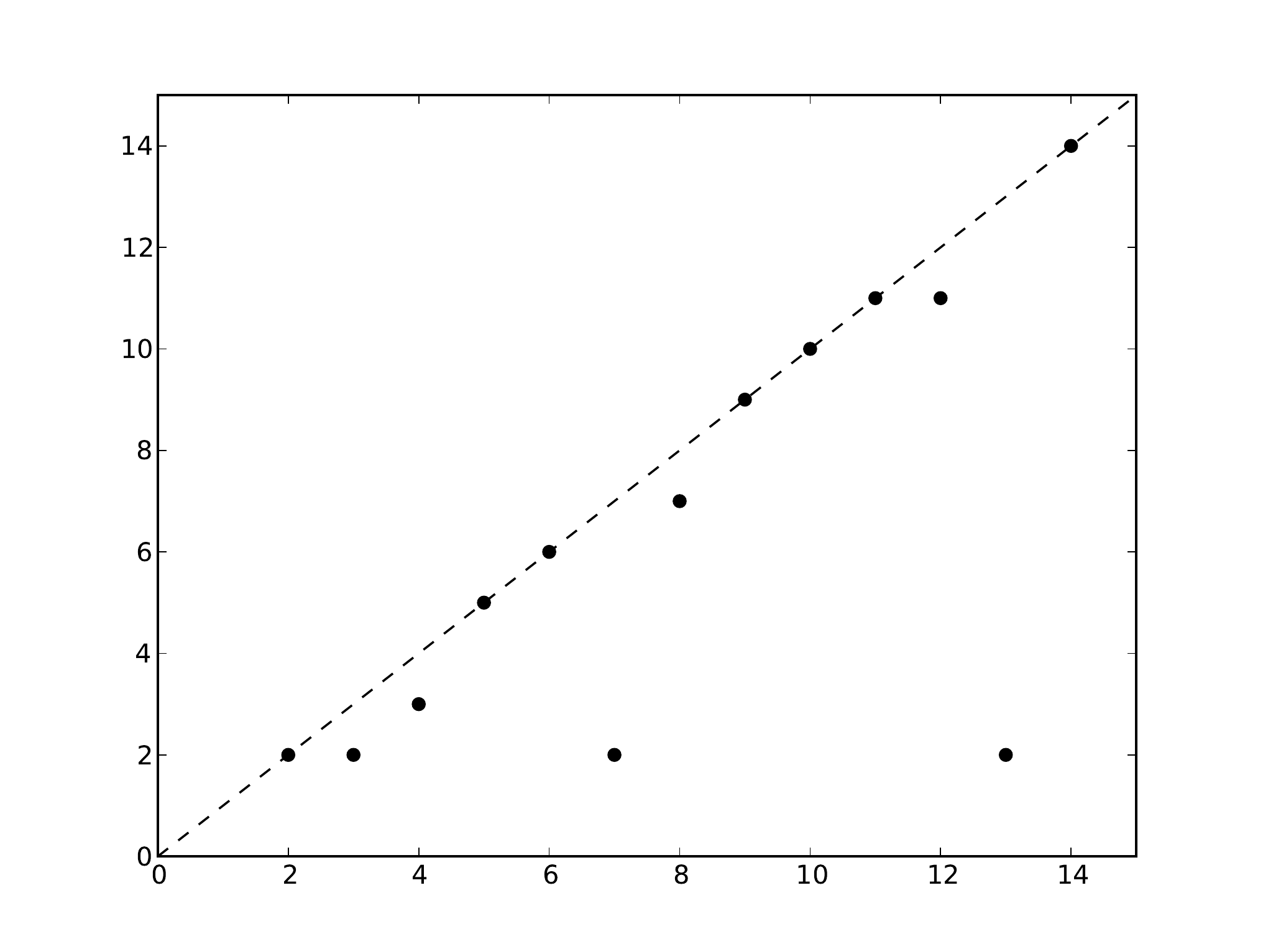}
                \caption{R154}
        \end{subfigure}
        \caption{\small A representative sample of rules. Horizontal axis displays the size of the automaton, vertical axis displays the required size of subsets to recover 90\% of the informational content. The dotted line corresponds to maximal A-complexity $\langle \Sigma \rangle_T=N$.}
\label{sampleadj}
\end{figure*}

A careful examination of the size of subsets required to reconstruct at least 90\% of the multi-information $M$ for all 88 inequivalent rules separately allows to single out some representative behaviours typified by the rules displayed in figure \ref{sampleadj} (the selection is arbitrary to a certain extent). The simplest behaviour is provided by rule 14, which belongs to Wolfram class II and exhibits a stable translating stationary configurational pattern (by \textit{configurational pattern} we shall always mean the spatial pattern obtained by evolving some initial configuration). In this case the number of coefficients to take into account is seen to be the same (here $\langle \Sigma \rangle_T=5$) whatever the size of the system, except for small sizes. This behaviour may be encountered in many rules, with some variations regarding the value of $\langle \Sigma \rangle_T$ or the length of the transient phase (being meant as the transient in terms of $N$ - recall that the time plays no role here since coefficients are averaged). Two interesting variations on this theme are to be found in rules 77 (reaching a stable inhomogeneous configuration and thus classified as class II) and 32 (reaching a stable homogeneous configuration and thus classified as class I). In the former case, $\langle \Sigma \rangle_T$ seems to get stabilized at $\langle \Sigma \rangle_T=4$, but jumps to $\langle \Sigma \rangle_T=3$ when $N$ increases from $N=8$ to $N=9$. This illustrates a weakness of our display of results, since $\langle \Sigma \rangle_T$ changes abruptly when, say, the first three coefficients contribute to 89\% of $M$, or when these same three coefficients reach 91\% of $M$. $\langle \Sigma \rangle_T$ would then jump from $\langle \Sigma \rangle_T=4$ to $\langle \Sigma \rangle_T=3$ although the change actually occurred almost smoothly. The case of rule 32 is more relevant to our purpose. There we oscillate between $\langle \Sigma \rangle_T=2$ and $\langle \Sigma \rangle_T=3$ depending on the parity of $N$. This may be explained by noting that there exists one initial configuration which does not converge to a stationary homogeneous pattern; namely, the alternate configuration $01010101010101$ gets replicated again and again over time except for a one-cell shift at each iteration. Such an initial pattern is however only possible for even values of $N$. It therefore happens that while the dynamics is rightly classified as class I for odd values, it should not be so for even ones (at least not for all initial configurations). It is therefore no longer unexpected to detect a size-dependent amount of A-complexity.

A completely different behaviour is exemplified by rule 90 already discussed above. Here $\langle \Sigma \rangle_T$ is close to its largest possible value $\langle \Sigma \rangle_T=N$ whatever the value of $N$. This is in accordance with the temporal behaviour of $C_k$'s observed in figure \ref{r90a}. Note the special case $N=8$, for which $\langle \Sigma \rangle_T$ drops to $\langle \Sigma \rangle_T=2$. A similar behaviour may be found in rule 106, which is perhaps even more convincing due to the lack of drop. Interestingly, while rules 90 and 106 present similarities, they are usually assigned to different Wolfram classes (R90 is class III while R106 is class IV). We shall come back to this question later on.

Both types of behaviour analyzed so far constitute extreme situations: in one case (rules 14, 77 and 32), $\langle \Sigma \rangle_T$ shows little or no dependance at all on the size of the system, while in the other (rules 90 and 106) $\langle \Sigma \rangle_T$ seems to grow linearly with $N$. The following rules do not display such unambiguous behaviours. Rule 105, for instance, oscillates between a ``basis line'' at $\langle \Sigma \rangle_T=2$ and an ``upper bound'' at $\langle \Sigma \rangle_T=N$, visiting intermediate values for some other values of $N$. Rule 15 behaves similarly, except that oscillations are sharper (intermediate values are never visited, at least for the range of sizes we were able to explore). Some other rules also exhibit what could be daringly (given the small range of sizes we were able to scan) characterized as sub-linear growth. Rule 22 illustrates this nicely, as well as rules 73, 110 and 54 (though in this latter case it is tempting to assert that $\langle \Sigma \rangle_T$ eventually gets stabilized at $\langle \Sigma \rangle_T=8$).

\begin{figure*}
        \centering
        \begin{subfigure}[b]{0.3\textwidth}
                \includegraphics[scale=0.25]{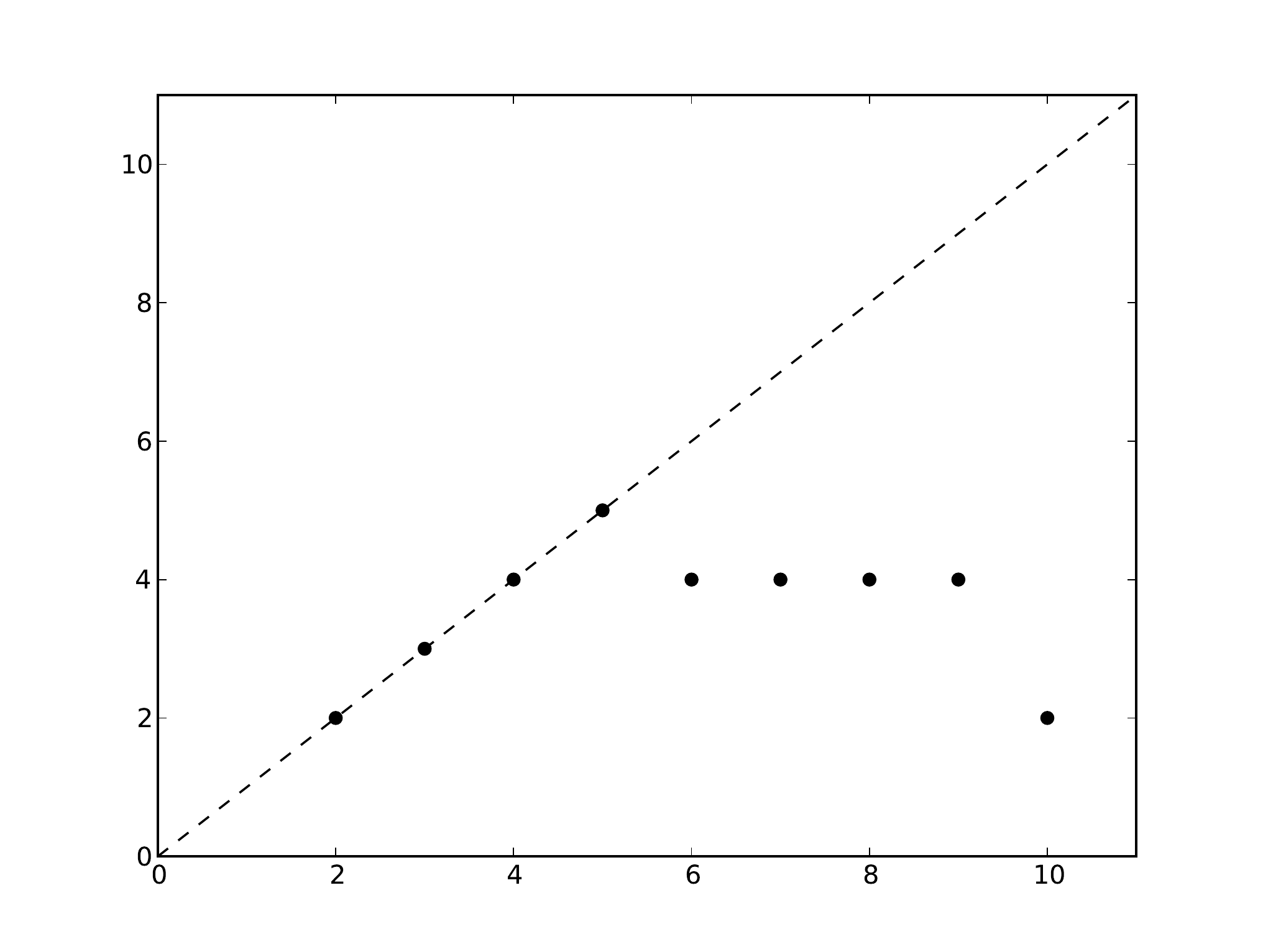}
                \caption{R14}
        \end{subfigure}
        \quad
        \begin{subfigure}[b]{0.3\textwidth}
                \includegraphics[scale=0.25]{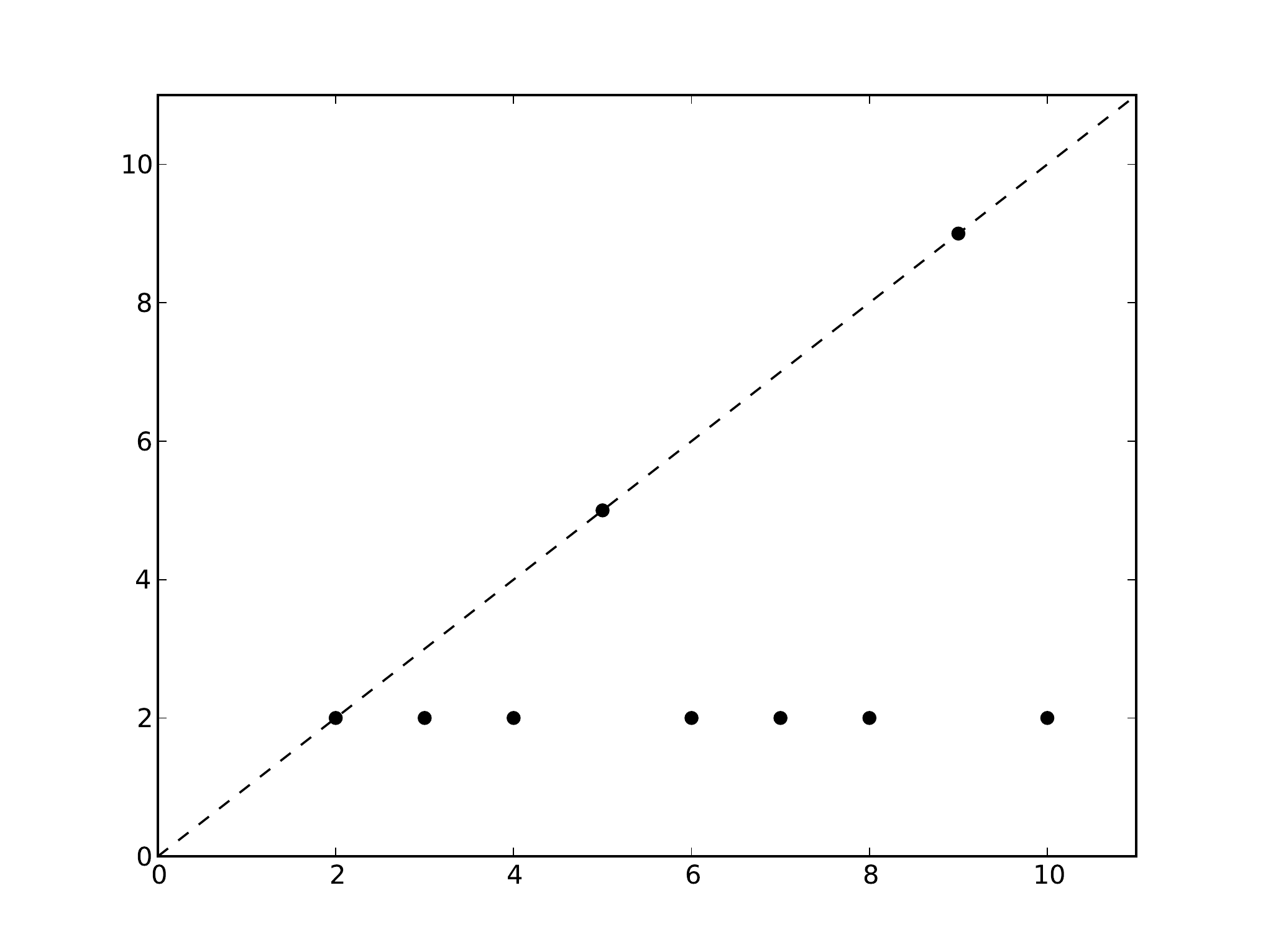}
                \caption{R15}
       \end{subfigure}
       \quad
        \begin{subfigure}[b]{0.3\textwidth}
                \includegraphics[scale=0.25]{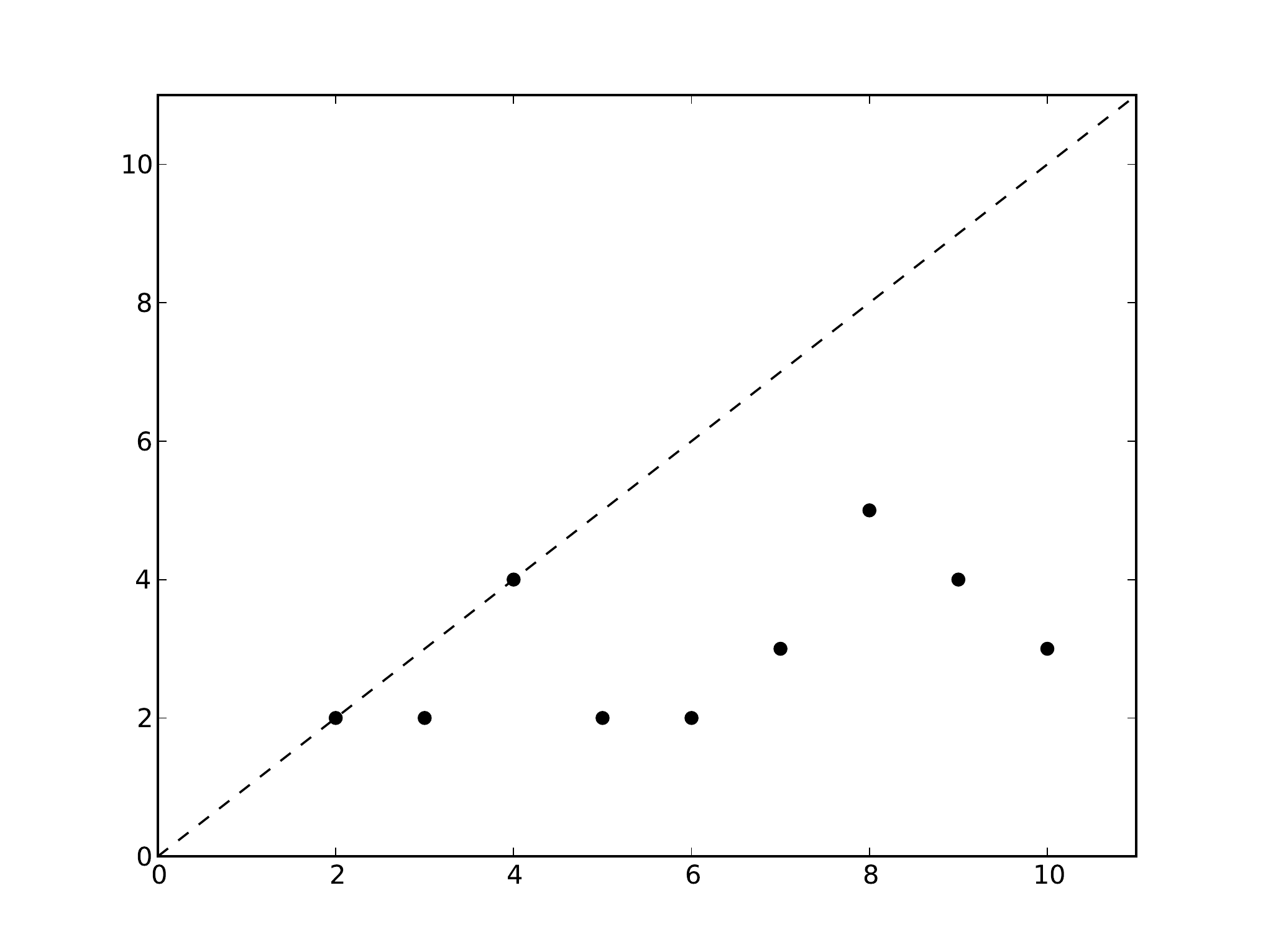}
                \caption{R22}
        \end{subfigure}

        \begin{subfigure}[b]{0.3\textwidth}
                \includegraphics[scale=0.25]{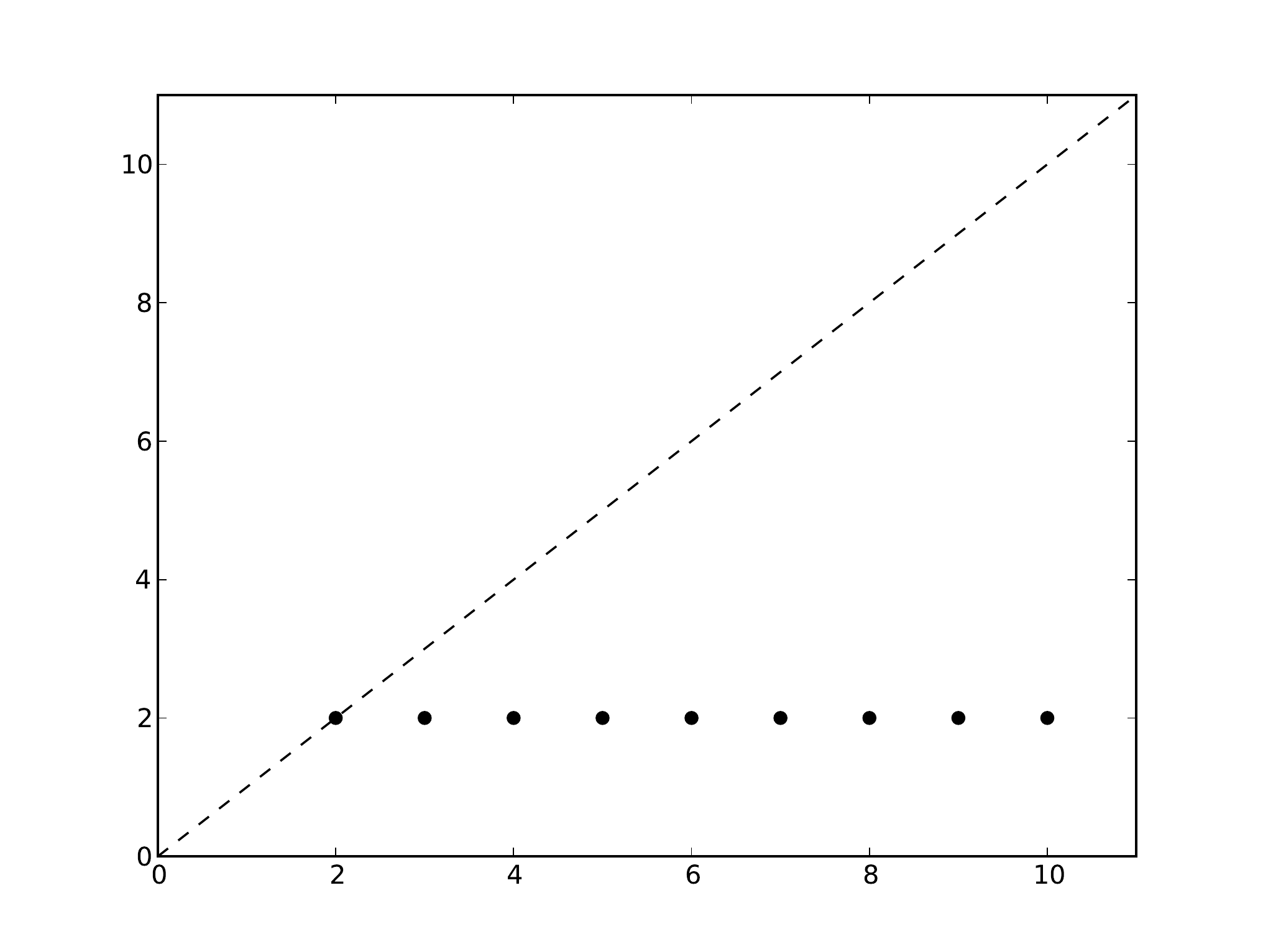}
                \caption{R32}
        \end{subfigure}
	\quad
	\begin{subfigure}[b]{0.3\textwidth}
                \includegraphics[scale=0.25]{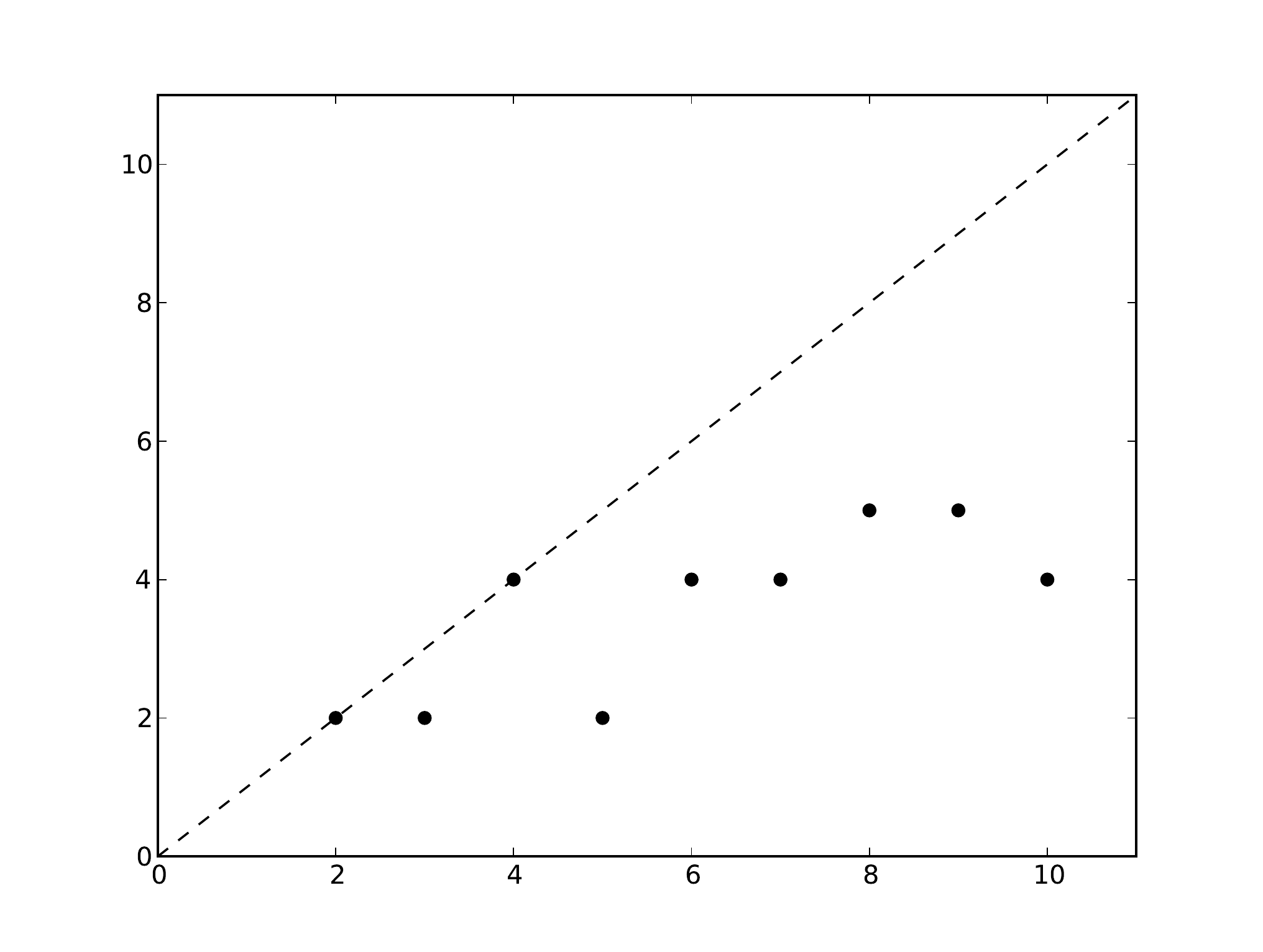}
                \caption{R54}
        \end{subfigure}
	\quad
        \begin{subfigure}[b]{0.3\textwidth}
                \includegraphics[scale=0.25]{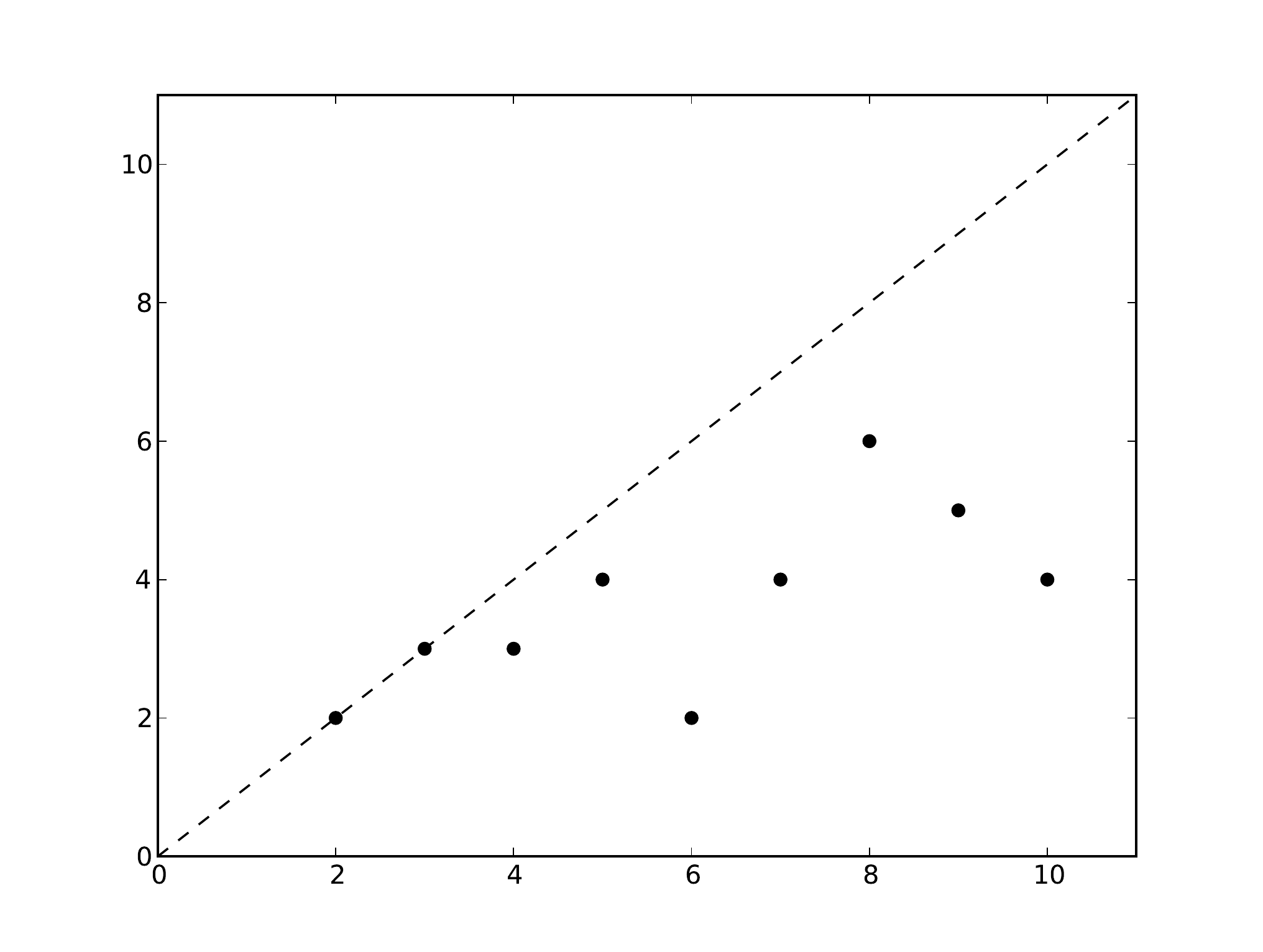}
                \caption{R73}
        \end{subfigure}

	\begin{subfigure}[b]{0.3\textwidth}
                \includegraphics[scale=0.25]{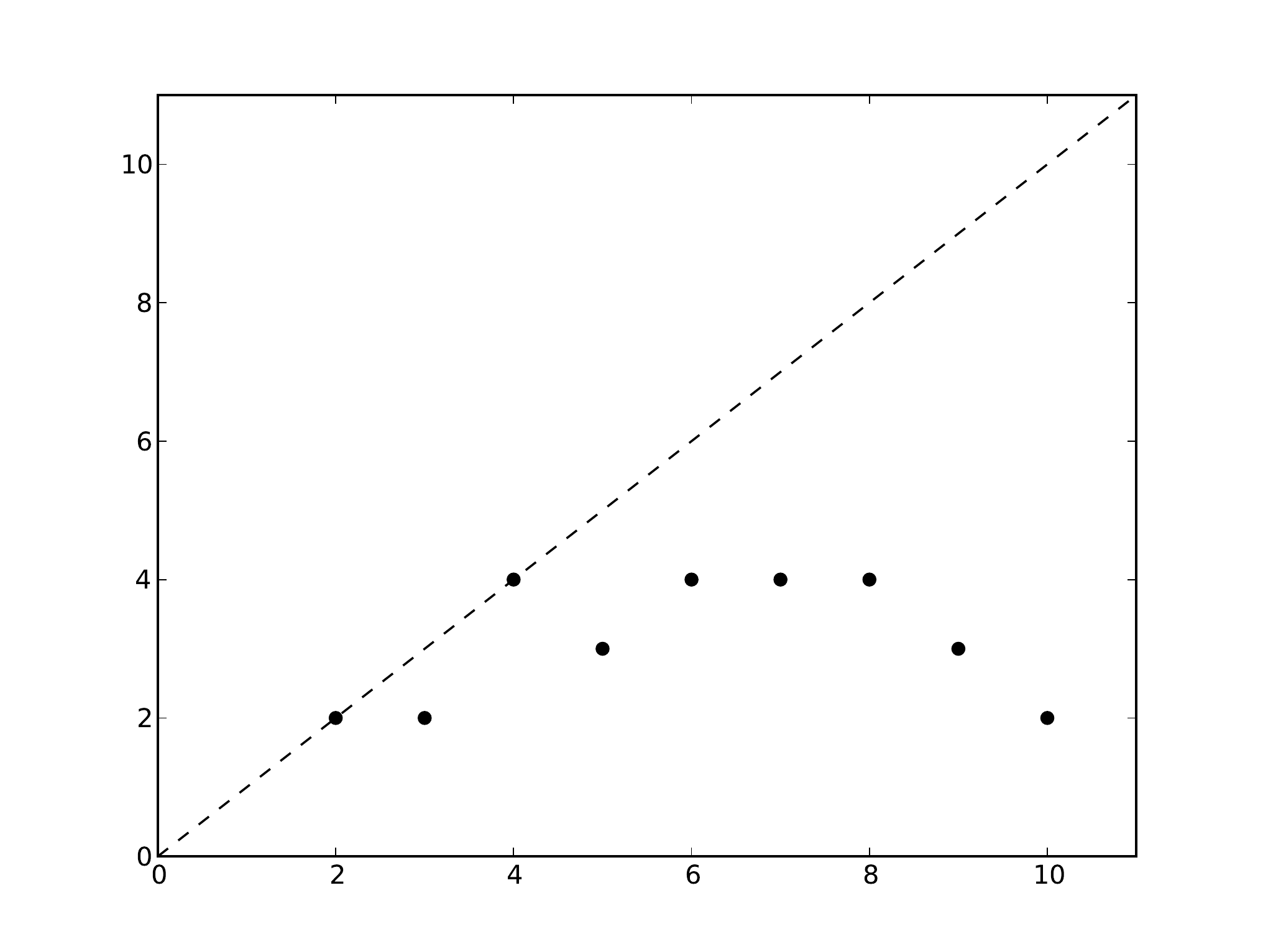}
                \caption{R77}
        \end{subfigure}
	\quad
        \begin{subfigure}[b]{0.3\textwidth}
                \includegraphics[scale=0.25]{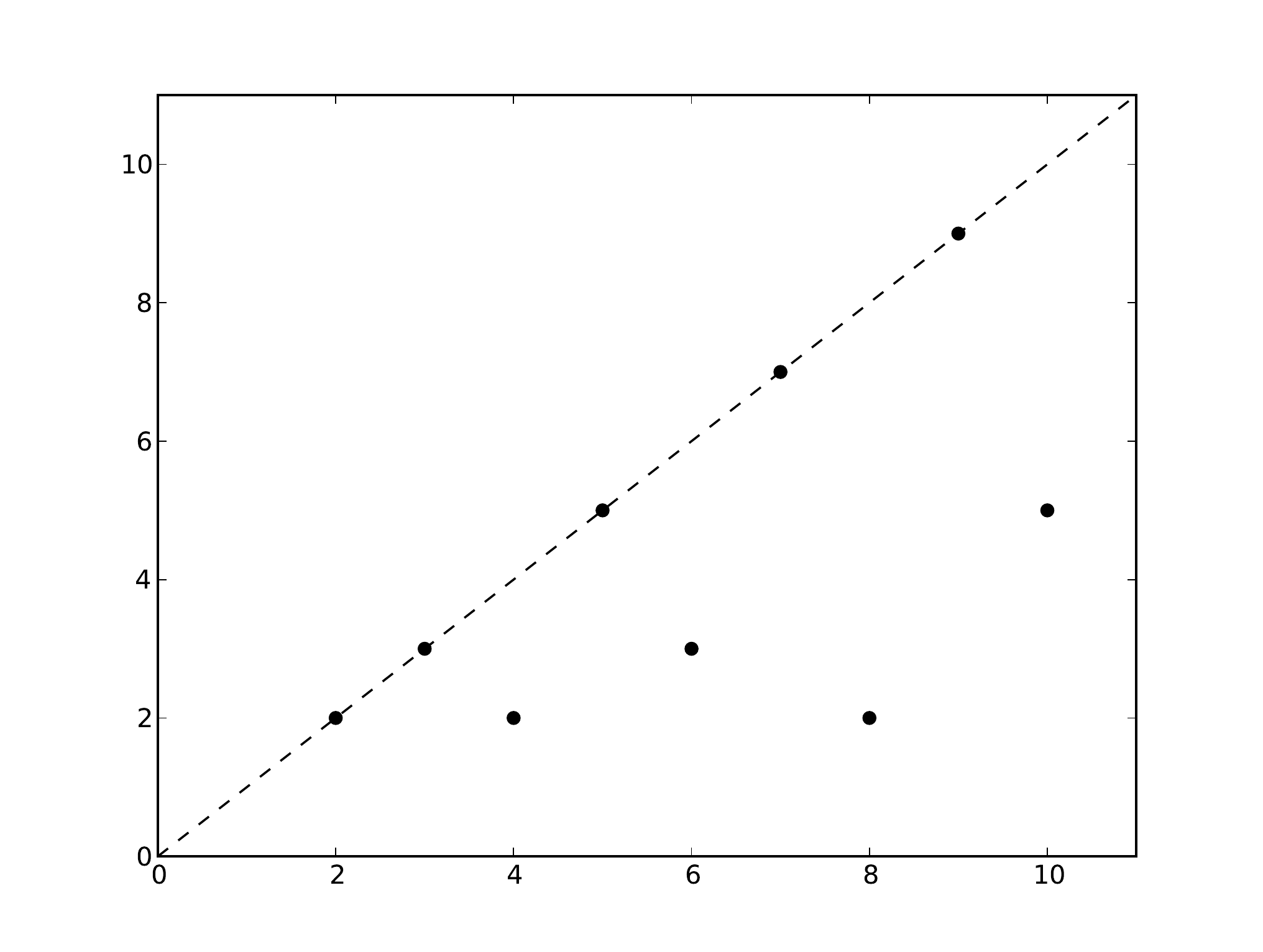}
                \caption{R90}
        \end{subfigure}
	\quad
	\begin{subfigure}[b]{0.3\textwidth}
                \includegraphics[scale=0.25]{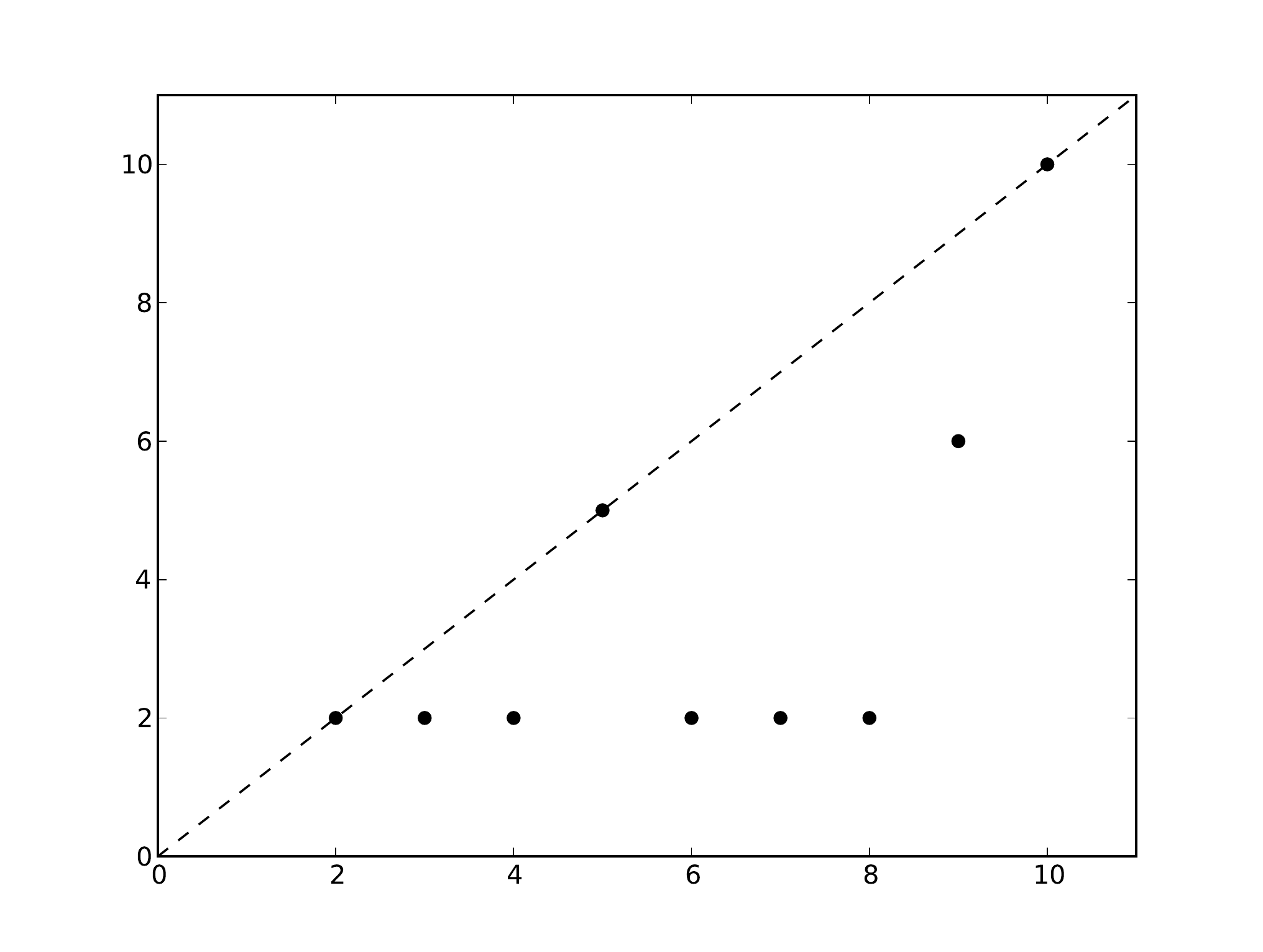}
                \caption{R105}
        \end{subfigure}

        \begin{subfigure}[b]{0.3\textwidth}
                \includegraphics[scale=0.25]{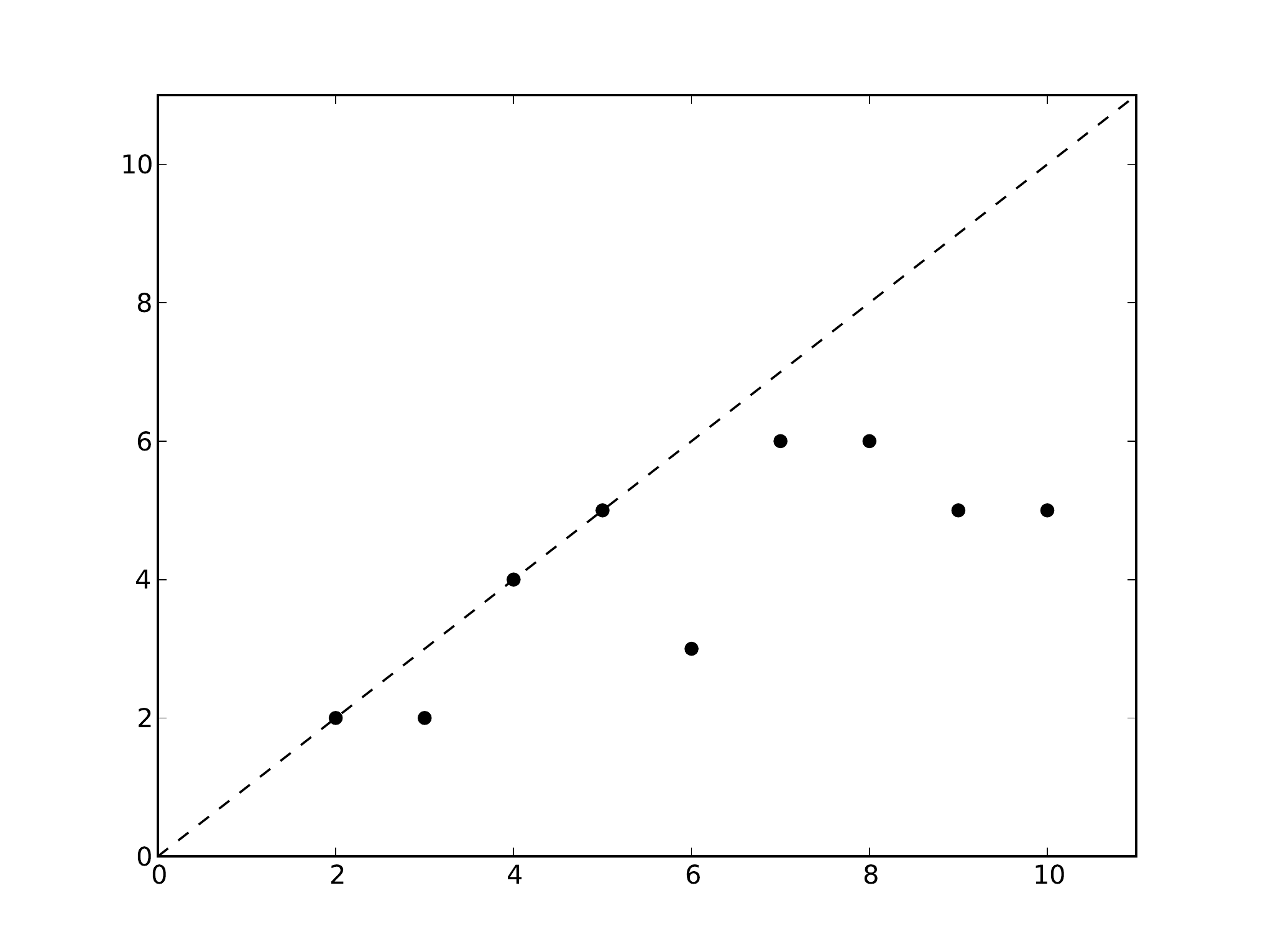}
                \caption{R106}
        \end{subfigure}
	\quad
	\begin{subfigure}[b]{0.3\textwidth}
                \includegraphics[scale=0.25]{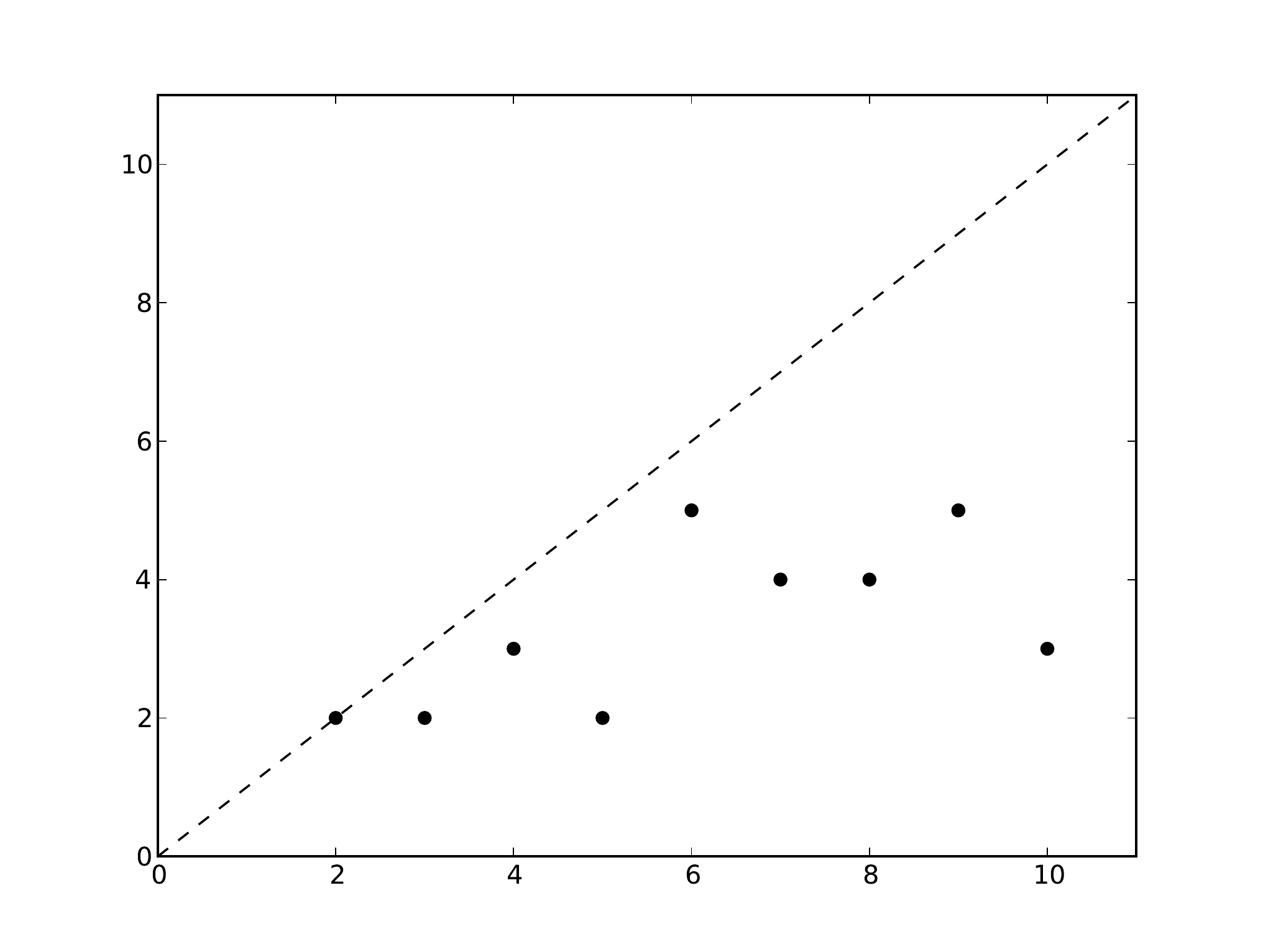}
                \caption{R110}
        \end{subfigure}
	\quad
        \begin{subfigure}[b]{0.3\textwidth}
                \includegraphics[scale=0.25]{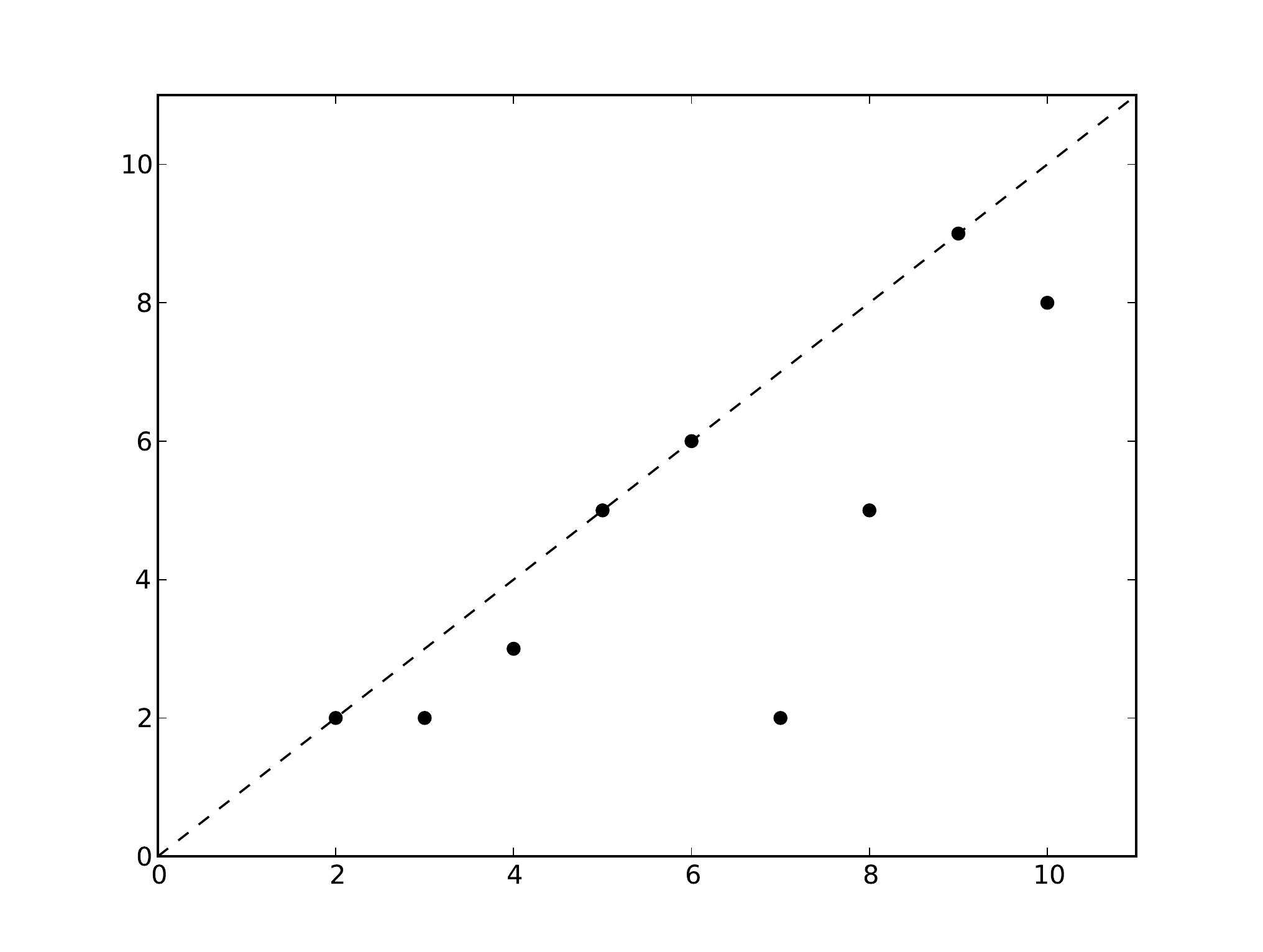}
                \caption{R154}
        \end{subfigure}
        \caption{\small The same picture as previously, now for the case where all subsets of a given size are taken as admissible. Only sizes up to $N=10$ are considered.}
\label{sampleall}
\end{figure*}

Before moving on to examine if and how these typical behaviours may be related to Wolfram's scheme of classification, let us have a side look on the situation where all subsets are considered instead of adjacent ones only. The counterpart of figure \ref{sampleadj} is displayed in figure \ref{sampleall}. It should first be noted that, as expected, the size of subsets to be considered in order to reconstruct a specified fraction of $M$ is almost always smaller when all subsets may be considered than when unconnected ones are to be discarded. Even if the results obtained in these two cases are somewhat different, the distinctive features underlined above are preserved. In particular, low A-complexity expressed in terms of adjacent subsets is coherent with low A-complexity expressed in terms of all subsets, and similarly for high A-complexity.

The primary purpose of this study was to put alongside the notion of complexity promoted by ME methods applied to Aristotle's principle and complexity as meant in Wolfram's classification. We should therefore examine whether, or not, some behaviours can be found to be common to all rules pertaining to a given Wolfram class. Table \ref{table1} lists all inequivalent rules and the class they belong to \cite{Martinez2013}.

\begin{table}
\centering
\begin{tabular}{ | c | p{7cm} |}
\hline
CLASS & RULES \\ \hline
I & 0, 8, 32, 40, 128, 136, 160, 168  \\ \hline
II & 1, 2, 3, 4, 5, 6, 7, 9, 10, 11, 12, 13, 14, 15, 19, 23, 24, 25, 26, 27, 28, 29, 33, 34, 35, 36, 37, 38, 42, 43, 44, 46, 50, 51, 56, 57, 58, 72, 73, 74, 76, 77, 78, 104, 108, 130, 131, 132, 133, 134, 138, 140, 142, 152, 154, 156, 162, 164, 170, 172, 178, 184, 200, 204, 232 \\ \hline
III & 18, 22, 30, 45, 60, 90, 105, 129, 146, 150, 161 \\ \hline
IV & 41, 54, 106, 110 \\
\hline
\end{tabular}

\caption{\small A compendium of the 88 inequivalent one-dimensional ECA}
\label{table1}
\end{table}

All eight rules in class I display the simple behaviour discussed above where $\langle \Sigma \rangle_T=const$. The value of $\langle \Sigma \rangle_T$ varies from one rule to the other. The tricky case of rule 32 has already been discussed above, and R160 is very similar. This homogeneity of behaviours is in agreement with the simplicity of configurational patterns converged to in the stationary regime.

At the other end of the spectrum, 11 rules belong to class III. All of them display linear or sub-linear growth, possibly with drops for certain values of $N$ (\textit{cf.} the discussion of R90 and R105 above). None of these rules shows the simple behaviour encountered in class I: here again, the ME approach is appropriate to catch the kind of complexity displayed by chaotic dynamics.

Things are no longer that clear when we come to considering classes II and IV. Class II regroups as much as 65 of the 88 inequivalent rules. At least 41 of them display the same simple type of behaviour already encountered in class I, which is fine since rules in class II are not expected to present a high level of complexity. Nonetheless, the remaining 24 rules behave in a way which is typical of class III. While this might suggest that classifications based on A-complexity on the one side and W-complexity on the other are definitively at odds with each other (which does not necessarily imply that one should be preferred to the other), it is also possible that the initial configuration plays an important role in this respect. We already met above a rule (R32) whose classification depended tightly on the initial configuration chosen, and, implicitly, on the size of the system. Another such instance is provided by rule 73, which is classified as class II due to the appearance of ``walls'' splitting the configurations into sub-configurations which, being of finite size, will necessarily repeat themselves, whence the attribution to class II. It may however happen that the inital configuration is chosen in such a way as to forbid the appearance of such separating walls, in which case the dynamics should better be classified as class III.

Class IV, finally, regroups only four inequivalent rules. Among these, three of which are shown above, one (R106) shows linear growth while another (R54) seems to converge towards a constant value of $\langle \Sigma \rangle_T=8$  as would dynamics in class I or II do (note however the large value of $\langle \Sigma \rangle_T$). Rule 110 lies somewhere in between. The fourth rule (R41) in class IV is similar to R110.

\section{VII. Concluding remarks}

It should be noted that we did not actually take full advantage of the ME machinery in this study, focussing instead on averaged quantities; the case of rule 110 serves as an illustration, since obviously the interplay of various coefficients shown in figure \ref{r110a} is much richer than the averaged values considered in most of our analysis. In spite of this, our results highlighted a tight relationship between A-complexity and W-complexity which is all the more remarkable. We cannot elude however that in some cases, namely in class II rules exhibiting very A-complex behaviour, considerable discrepancies arose between these two schemes. In our opinion this may be interpreted in two different ways. Firstly, it might be that our study should indeed be refined by looking at more subtle quantities than averaged ones. Nonetheless, we have met several cases (\textit{e.g.} rules 32, 73, 160) where unexpected A-complexity could be explained by misattributions to such-and-such a class due to unsufficient attention paid to particular initial conditions. A great strength of our probabilistic approach is that we cannot be fooled by such effects since all possible initial configurations are considered in our framework. Moreover, if one wishes, it allows a separate treatment of these rogue configurations by simply assigning them probability zero. The issue lies in the determination of these special initial conditions,  which would require a considerable amount of work.

Assuming all ambivalent rules may indeed be explained by an adequate splitting of initial configurations (which in itself would shed some light on the interplay between dynamics, initial conditions, and complexity of behaviour), it is very tempting to sketch the following global picture. \textit{1) In automata converging to a stable or periodic configurational pattern the knowledge of subsets of some finite size is sufficient to reconstruct accurately (here up to a 10\% error) the informational content of the system. 2) In chaotic automata the size of these subsets grows linearly with size, meaning that any inference based on small subsystems yields intrinsically flawed results. 3) Complex systems would then lie somewhere in between, perhaps exhibiting sub-linear growth of the required subsets}. In other words, W-complexity corresponds to an intermediate regime of A-complexity, quite akin to Langton's edge of chaos which is therefore recovered starting from a completely different vantage point.

The behaviour encountered in classes I and II is somewhat reminiscent of the situation prevailing in classical kinetic theory, where the BBGKY hierarchy of equations may be truncated without too much harm after two steps for most gaseous or fluid systems of interest, neglecting in a sense the contribution of higher-order reduced densities. Pushing further this analogy, this would suggest that such systems cannot be characterized as ``complex'' in the sense investigated here. This comparison is however made very conjectural by the fact that in the present study the criterion for truncation is provided by the informational content of the system, which is not the case in the context of kinetic theory where such a criterion is not so clearly stated. Actually, our criterion is rather arbitrary; indeed, reconstructing its informational content is but a first step towards the understanding of a system since there is no one-to-one relationship between probability distribution and multi-information.

This should remind us that the ME method employed is susceptible of several refinements, without even mentioning the fact that the cellular automata studied here are a very specific kind of system. As we already mentioned, different types of observational constraints may be used in the reconstruction of the probability density, and we see no reason why the $C_k$ coefficients could not be diversified accordingly, perhaps giving rise to tractable analytical expressions. We also emphasized that the selection of the subsets to take into account deserved careful attention. Lastly, it is unfortunate that these ME models are so difficult to handle analytically and computationally so demanding, precluding the exploration of larger systems; advances on the theoretical as well as on the numerical side are therefore mandatory if one wishes to gain insight into the underlying physics.

\begin{acknowledgments}
\section{Acknowledgments}
The authors would like to thank Joris Borgdorff, Christophe Charpilloz, Raphael Conradin, Alexandre Dupuis and  Anton Golub for their helpful advices and comments. The research leading to these results has received funding from the European Union Seventh Framework Programme (FP7/2007-2013) under grant agreement 317534 (Sophocles).
\end{acknowledgments}

\bibliography{References}

\end{document}